\documentclass[letterpaper,dvipsnames]{article}
\pdfoutput=1
\usepackage{jheppub}
\usepackage{amsmath}
\usepackage{graphicx}

\usepackage{array}
\usepackage{ulem}
\usepackage{cancel}
\usepackage{xcolor}

\usepackage{slashed}
\usepackage{afterpage}
\usepackage{appendix}
\usepackage{multirow}
\usepackage{float}
\usepackage{mathtools}
\usepackage{physics}
\mathtoolsset{centercolon}
\usepackage{soul}
\usepackage{ulem}
\usepackage{multirow}

\DeclareMathOperator{\arcsinh}{arcsinh}
\DeclareMathOperator{\arctanh}{arctanh}
\DeclareMathOperator{\arcsech}{arcsech}

\newcommand{\uv}{{\rm uv}}
\newcommand{\ir}{{\rm ir}}
\newcommand{\UV}{{\rm UV}}
\newcommand{\IR}{{\rm IR}}

\newcommand{\Hsigma}{H_\sigma}

\title{Inflation with a Growing Fifth Dimension}

\author{Rashmish K.~Mishra,}
\author{Michael Nee,}
\author{and Lisa Randall}
\affiliation{Jefferson Physical Laboratory, Harvard University,\\ 17 Oxford Street, Cambridge, MA, 02139, USA}

\abstract
{
Inflation generally assumes a field with nonzero potential that leads to inflationary expansion happening at arbitrarily early times. We demonstrate potentially observable consequences of inflation with a finite initial time in a model in five-dimensional warped anti-de Sitter space, with both a UV and an IR brane present during inflation. Considering an inflaton with an approximately flat potential localized on the UV brane, we derive the resulting brane motion in the bulk and the 4D effective action describing the dynamics. A concrete model allows us to evaluate possible consequences of a starting point of inflation. The background evolution is driven by the fast roll of the radion at early times and the slow roll of the inflaton at late times. We find that the action has the form of a two-field hyperbolic inflation model, the two fields being the radion and the inflaton, both of which have a time-dependent background solution. This setup is holographically dual to an inflaton coupled to a strongly coupled confining sector in which the ratio of the confinement scale to the 4D Planck scale evolves cosmologically. Focusing on the period when the equation of state becomes that of inflation, we find that the presence of the IR brane leads to deviations from the approximate de Sitter background in addition to those from the slow-roll parameters of the inflaton potential. We quantify the effect of the presence of the IR brane on the two point function of the adiabatic scalar perturbations and tensor perturbations. The dominant deviations occur at large scales: the adiabatic power spectrum has a blue tilt, while the tensor power spectrum shows oscillatory features. We present numerical fits to the shape of the adiabatic power spectrum, and discuss the implications for cosmic microwave background (CMB) analysis.
}

\begin{document}

\maketitle

\section{Introduction}

Time-dependent solutions in field theory are relevant to early Universe cosmology, most notably for cosmological inflation. While many models of inflation are constructed in a purely 4D theory, additional compact dimensions, which are ubiquitous in string theory, could introduce additional time-dependence. In principle, for sufficiently specific predictions, this could allow for tests that distinguish higher-dimensional inflation from the purely 4D counterpart. Inflationary models with static extra dimensions have been well-studied in the past~\cite{Lukas:1998qs, Lukas:1999yn,Arkani-Hamed:1999hav, Arkani-Hamed:1999fet, Nihei:1999mt, Flanagan:1999dc, Maartens:1999hf, Kaloper:1999sm,Jones:2002cv}; in this work we extend this idea to allow for time dependence in the extra dimensions. Refs.~\cite{Anchordoqui:2022svl, Anchordoqui:2023etp,  Anchordoqui:2024amx, Antoniadis:2025pet, Hirose:2025pzm} considered flat extra dimensions, but in this work we focus on a warped extra dimension.

Five-dimensional warped geometries consisting of bulk anti-de Sitter (AdS) space between a UV and an IR brane provide an interesting framework for constructing such time-dependent solutions and investigating possible imprints on cosmological observables. Using mismatched branes, ref.~\cite{Karch:2020iit} proposed such time-dependent solutions in which the UV/IR brane locations are not static due to the brane tensions 
deviating from their tuned values for the flat RS solution. Ref.~\cite{Karch:2020iit} also constructed the low-energy 4D action corresponding to such geometries, and showed that 4D gravity with one additional scalar (the radion), which has a time-dependent background solution, is the appropriate low energy description of such dynamics. 

Ref.~\cite{Karch:2020iit} considered only fixed brane tensions, but an intriguing possibility is that the tensions of the UV/IR branes change dynamically over their cosmological history due to the dynamics of brane-localized fields.  
In the 5D picture, this can trigger a time evolution of the system that eventually settles into a new static solution. In this work, we study the cosmological implications of brane-localized inflationary dynamics.

Our goal is to better understand this scenario and determine whether there are observable deviations from a purely 4D theory. In particular we consider an inflaton localized on the UV brane, where the UV brane location is a function of time driven by the localized inflaton potential and the flat IR brane is fixed.\footnote{The motion of branes is a coordinate-dependent statement. Here we are implicitly assuming a Minkowski slicing of the extra dimension. This is further discussed in app.~\ref{app:dS-slicing}.} The 4D effective theory then consists of a 4D graviton and two scalars, an inflaton and a radion, with coupled dynamics. The effective action has a similar form as hyperbolic two-field inflation models~\cite{DiMarco:2002eb, DiMarco:2005nq, Lalak:2007vi, Tolley:2009fg, Brown:2017osf, Mizuno:2017idt, Garcia-Saenz:2018ifx, Garcia-Saenz:2019njm, Bjorkmo:2019aev, Fumagalli:2019noh}. We find that due to the presence of an IR brane, the effective 4D metric deviates from pure de Sitter (dS) at early times (in addition to the deviations from the non-flatness of the inflaton potential). We quantify the effect of this deviation on the scalar and tensor power spectra and comment on the non-Gaussianities. As we will see, towards the start of inflation the branes are close to each other and the adiabatic fluctuations are not generated from the inflaton, but rather from the radion.

A novel feature of the dynamics here is that there is a starting point for inflation. This dynamics, for suitable parameters, can yield visible effects in the cosmic microwave background (CMB). This is to be contrasted with the usual inflationary models for which it is assumed that inflation started in the asymptotic past, which allows restricting to homogeneous background solutions since any inhomogeneities decay away. Furthermore, in usual inflationary models, all $k$ modes are deeply subhorizon in the asymptotic past, which justifies using the Bunch-Davies vacuum as the quantum state. In the scenario considered here, small $k$ modes experience fewer e-folds before exiting the horizon. This means that initial inhomogeneities have not had sufficient time to decay, and the quantum fluctuations are sensitive to the choice of initial conditions. We choose parameters such that the CMB modes do experience sufficient number of e-folds between exiting the horizon and reentry to address the horizon and flatness problems.

Holography offers a dual interpretation of the inflationary dynamics presented here. In the dual picture, the UV degrees of freedom are elementary, and the bulk encodes dynamics of a strongly coupled conformal field theory (CFT)~\cite{Arkani-Hamed:2000ijo, Rattazzi:2000hs}. The IR brane signals that the CFT confines in the IR. The dynamics here then corresponds to a strongly coupled confining sector coupled to an inflaton. The state of the theory that we study in this work is that of a confining sector with a time-dependent ratio of the confinement scale to the Planck scale. While such a state sounds exotic from a purely field-theoretic viewpoint, it arises naturally in the bulk description, coming from the detuning and mismatch of brane tensions. It is natural to expect brane-tensions to be time-dependent, since they are intertwined with the dynamics localized on the branes. The 5D geometric setup naturally gives access to interesting time-dependent states of the theory and allows making concrete quantitative predictions for these scenarios.

There is significant literature on the presence of additional degrees of freedom present during inflation, and most of the work has focused on weakly coupled scenarios. Although there is some literature on considering a strongly coupled sector coupled to an inflaton (e.g. see refs.~\cite{Green:2013rd, Baumgart:2021ptt} for a purely CFT based approach, ref.~\cite{Aoki:2023tjm} for an approach using spectral functions, ref.~\cite{Hubisz:2024xnj} using spectral functions and holography, refs.~\cite{Pimentel:2025rds, Yang:2025apy} using unparticles, and ref.~\cite{Chakraborty:2025myb} in the context of light compact scalars), most of these works have focused on the CFT in the gapless phase. Our work instead focuses on a strongly coupled sector in the gapped phase (see also~\cite{Kumar:2018jxz, Kumar:2025anx}), and leads to novel signals in the inflationary observables. Our work therefore provides an important and well-motivated theoretical alternative to the existing literature.

The rest of the paper is organized as follows. In sec.~\ref{sec:EFT-of-mismatched-branes} we consider a 5D warped geometry with a detuned UV brane, derive the brane motion from the detuning and derive the resulting 4D effective field theory (EFT), most of our steps being a review of~\cite{Karch:2020iit}. In sec.~\ref{sec:UV-localized-inflaton} we consider the detuning of the UV brane to arise from a slowly rolling inflaton and derive the 4D action involving the inflaton and the radion. We calculate the equation of state and the background solutions to first order in the slow-roll parameters. We also identify the regime of validity of our description and the start/end of inflation. In sec.~\ref{sec:linear-fluctuations} we derive the equations governing linear scalar fluctuations, identify the mass of the fluctuations, and decompose them into adiabatic and entropy components. We then solve for the mode function numerically and analytically for our choice of vacuum. In sec.~\ref{sec:adiabatic-and-cmb} we numerically calculate the power spectrum of the adiabatic scalar fluctuations and identify the effect on the amplitude and tilt from the presence of the IR brane. We also present fits to the numerical results, and quantify the predictions for the CMB power spectrum. In sec.~\ref{sec:isocurvature-and-NG} we comment on the isocurvature and the non-Gaussianities that can arise in the model. In sec.~\ref{sec:linear-fluctuations-tensor} we consider tensor fluctuations and analytically calculate the power spectrum of the tensor modes. Finally, we summarize the main results and conclude in sec.~\ref{sec:conclusion}.

\section{EFT of Mismatched Branes}
\label{sec:EFT-of-mismatched-branes}

In this section we consider a 5D warped geometry between a UV and an IR brane, where the branes are not stabilized and do not have tuned brane tensions. Most of our steps are a review of ref.~\cite{Karch:2020iit}. We first determine the solutions to the equations of motion in 5D and use these to derive the resulting 4D EFT.

Our starting point is the 5D action
\begin{align}
    S_5 &= 2M_5^3 \int d^5 x \sqrt{\left|G\right|} \left(R_{5} 
    +\frac{12}{L^2}
    \right) + \sum_{i=\uv, \ir} \int d^4 x \sqrt{\left|g_i\right|}\left(\mathcal{L}_i + T_i\right)\:,
    \label{eq:5D_action-general}
\end{align}
where $M_5$ is the 5D Planck scale, $L$ is the AdS radius, 
$T_i$ are the brane localized tensions and $\mathcal{L}_i$ are brane localized Lagrangians of matter fields, at the $i = \UV, \IR$ branes. The flat sliced RS solution~\cite{Randall:1999ee, Randall:1999vf} is obtained when the tensions are tuned to $T_\UV = 12M_5^3/L , T_\IR = - 12 M_5^3/L $ and there are no additional contributions to the tension from the matter fields on the brane. If the two branes are detuned in a correlated way, one obtains the AdS or dS slicing~\cite{Karch:2000ct}. Ref.~\cite{Karch:2020iit} generalized the setup by considering the case when the detunings of the two branes are not correlated, dubbed \textit{mismatched} branes. The resulting 5D geometry can be consistently sliced but then the radion has a time-dependent background. Ref.~\cite{Karch:2020iit} observed that even if the branes are mismatched, the resulting solution can be written as 
\begin{align}
    d s^2 = \frac{L^2}{z^2}\left(g_{\mu\nu}dx^\mu d x^\nu+ dz^2\right)\, \qquad z_\uv \leq z \leq z_\ir\:,
    \label{eq:5D_metric-general}
\end{align}
where $g_{\mu\nu}$ is the 4D metric, $L$ is the AdS radius and $z_\uv, z_\ir$ are the locations of the UV/IR branes, which are crucially now time-dependent. The exact time dependence depends on the detuning of the respective tensions and can be obtained by solving the Israel junction conditions in 5D. The time dependence in $z_\uv, z_\ir$ makes the spacetime described by the metric in eq.~\eqref{eq:5D_metric-general} time-dependent. 
In general, the explicit time dependence of $z_\uv, z_\ir$ depends on the choice of coordinates, also called the choice of slicing. Here we will work with the coordinates defined in eq.~\eqref{eq:5D_metric-general}, referred as the flat slicing (see app.~\ref{app:dS-slicing} for another choice of coordinates, the de Sitter slicing). Working with the flat slicing has the advantage that the 4D metric in conformal time coordinate is very straightforward to obtain, as we will see in sec.~\ref{subsec:4D-picture}.
We denote the 5D coordinates in eq.~\eqref{eq:5D_metric-general} by $(\tau, x, z)$ and the proper time on the UV brane by $\lambda$. We use $\eta$ to denote the conformal time in the 4D EFT, and $t$ to denote the usual time coordinate in 4D EFT, where $g_{00} = -1$. Derivatives w.r.t. $t$ will be denoted by a dot, and w.r.t. $\eta$ will be denoted by a prime. 5D indices will be denoted by upper case roman letters, 4D indices by greek letters and 3D spatial indices by lower case roman letters.

We will first consider a constant detuning of the UV brane tension. The IR brane tension will be assumed to be at its tuned value, so that it does not move in our assumed frame, although this assumption can be easily relaxed. In the following subsections we derive the explicit time dependence of $z_\uv$ and the 4D EFT describing the dynamics.

\subsection{5D picture}
\label{subsec:5D-pic}

In the 5D picture, the time dependence of the background solution comes from the motion of the branes in the bulk (here we are taking the UV brane tension to be constant). Working in flat slicing (i.e. the bulk metric given by eq.~\eqref{eq:5D_metric-general}), the UV brane moves as a result of the localized tension on the UV brane from the inflaton potential, while the IR brane is at rest.\footnote{In de Sitter slicing, the UV brane would be at a constant location while the IR brane would move. The dynamics is of course independent of the choice of slicing. See app.~\ref{app:dS-slicing} for more details.} The motion of the UV brane can be derived by the Israel junction conditions~\cite{Kraus:1999it}. Working in the coordinates~\eqref{eq:5D_metric-general}, the brane world-volume can be parameterized by $(\tau(\lambda), z_\uv(\lambda))$, where $\lambda$ is the proper time on the brane. The brane velocity vector $u^{M}$ (which satisfies $u^M u_M = -1$), and the normal vector $n^M$ (which satisfies $n^Mu_M = 0, n^M n_M = 1$) are given respectively as
\begin{align}
    u^{M} &= \left(\frac{\dd \tau}{\dd \lambda}, 0,0,0,\frac{\dd z_\uv}{\dd \lambda}\right) = 
    \left(
    \sqrt{(\partial_\lambda z_\uv)^2 + (z/L)^2},0,0,0,\partial_\lambda z_\uv
    \right)\:,
    \label{eq:v-vector}
    \\
    n^M &= \left(\partial_\lambda z_\uv,0,0,0, \sqrt{(\partial_\lambda z_\uv)^2 + (z/L)^2} \right)\:.
\end{align}

The line element squared on the moving UV brane is given as
\begin{align}
    \dd s^2_\text{uv}&=-\frac{L^2}{z_\uv^2}\left(1-\left(\frac{\dd z_\uv}{\dd \tau}\right)^2\right)\dd \tau^2 + \frac{L^2}{z_\uv^2}\dd x^2\:,
\end{align}
from which the induced metric $\gamma_{\mu\nu}$ can be read off. The extrinsic curvature is given by
\begin{align}
    K_{\mu\nu} &=  \nabla_\mu n_\nu  \:,
\end{align}
in terms of which the junction condition is
\begin{align}
    K_{\mu\nu}^+ - K_{\mu\nu}^- &= -\frac{1}{4M_5^3}\left(T_{\mu\nu} - \frac13 T_\rho^\rho \gamma_{\mu\nu}\right)\:,
    \label{eq:junction_1}
\end{align}
where $+(-)$ refer to the $z>z_\uv (z<z_\uv)$ region. Here the only contribution comes from $z>z_\uv$, i.e. $K^-_{\mu\nu} = 0$. The localized energy momentum tensor is given as $T_{\mu\nu} = -(T_c + \delta T_\uv)\gamma_{\mu\nu}$, where $\delta T_\uv$ is the detuning from the critical value $T_c = 12 M_5^3/L$. As the spatial derivatives of $n_M$ vanish, it is simplest to look at the spatial components of the extrinsic curvature, which are given by
\begin{align}
    K_{ij} &= - \Gamma^M_{ij} n_M = \frac{\gamma_{ij}}{L}\sqrt{1 + \frac{L^2 (\partial_\lambda z_\uv)^2}{z_\uv^2} } \, ,
\end{align}
where $\Gamma^L_{MN}$ are the Christoffel symbols derived from the 5D metric~\eqref{eq:5D_metric-general}. This leads to the junction condition from the $i, j$ component of equation~\eqref{eq:junction_1}\footnote{We note that the $00$ component is different than~\eqref{eq:JunctionEquation} but has the same solution for $z_\uv$.}
\begin{align}
	\sqrt{1 + \frac{L^2 (\partial_\lambda z_\uv)^2}{z_\uv^2} }
    &= 1+ \frac{\delta T_\uv}{T_c}\:.
    \label{eq:JunctionEquation}
\end{align}
The junction condition can be solved to give
\begin{align}
	&z_\uv(\lambda) = z_0 e^{\pm\frac{\lambda}{L} \sqrt{\frac{\delta T_\uv}{T_c} \left(2+\frac{\delta T_\uv}{T_c}\right)}}\: ,
    \label{eq:zuv-sol}
\end{align}
for some initial value $z_\uv(0) = z_0$. We note that both signs in the exponent are valid solutions. However, we will take the negative sign of the exponent so that $z_\uv$ decreases with time. This choice corresponds to the de Sitter metric on the brane, with a growing scale factor as time increases. The second solution with a positive exponent corresponds to a contracting Universe described by a time-reversed de Sitter metric.

Using eq.~\eqref{eq:v-vector}, we can calculate the relation between $\tau$ and $\lambda$ (using $\tau=0$ when $\lambda = 0$):
\begin{align}
    \exp\left(-\frac{\lambda}{L}\sqrt{\frac{\delta T_\uv}{T_c}\left(2+\frac{\delta T_\uv}{T_c}\right)}\right) = 1-\frac{\tau}{z_0\,(1+\delta T_\uv/T_c)}\sqrt{\frac{\delta T_\uv}{T_c}\left(2+\frac{\delta T_\uv}{T_c}\right)}\:,
\end{align}
using which one can obtain $z_\uv(\tau)$.
In the following section we will use this solution to match to the 4D effective theory which comes from integrating over $z$. The 4D Friedmann equations of motion match the 5D equations of motion to leading order in $\delta T_\uv/T_c$~\cite{Cline:1999ts, Csaki:1999jh, Binetruy:1999hy, Csaki:1999mp}, so we keep terms to leading order in $\delta T_\uv$ and get
\begin{align} \label{eq:zuv5D}
    z_\uv(\tau) &= z_0 - \gamma\, \tau\:,\qquad
    \gamma = \left(\frac{2\delta T_\uv}{T_c}\right)^{1/2}\:.
\end{align}

\subsection{4D picture}
\label{subsec:4D-picture}

To derive the 4D EFT, we follow the strategy outlined in ref.~\cite{Karch:2020iit}. The main observation is that due to the time dependent brane locations, the volume of the extra dimension is also time dependent. This time-dependence is captured by the radion $\phi$, which  parameterizes the 5D volume in the EFT and is coupled to the 4D Ricci scalar $R_4$. The radion and graviton are the only light fields in the EFT, as we integrate out all the Kaluza-Klein (KK) modes of the 5D graviton.\footnote{In the next sections we will also incorporate a light inflaton into the EFT.}

To see this we integrate the 5D Einstein-Hilbert term over the extra dimension, keeping just the 4D Ricci scalar, $R_4$, after the integration:
\begin{align}
    \mathcal{L}_{\rm grav} &= 2M_5^3 \int_{z_\uv}^{z_\ir} dz \sqrt{-G} R_5 
    \:\supset\: M_5^3L \left(  \frac{L^2}{z_\uv^2} -\frac{L^2}{z_\ir^2} \right) \sqrt{-g} R_4 \: .
    \label{eq:dim_reduction}
\end{align}
This has the form $\mathcal{L}_{\rm grav} = \frac12 \sqrt{-g}  M_4^2 \phi R_4$, where:
\begin{align}
    M_4^2 &= 2M_5^3 L \, , \:\:
    \phi = \left( \frac{L^2}{z_\uv^2} -\frac{L^2}{z_\ir^2} \right) \, .
    \label{eq:phidef}
\end{align}
If the branes are moving, at least one of $z_\uv$ and $z_\ir$ is time-dependent and the effective Planck scale in the 4D theory, $M_4\phi^{1/2}$, is time-dependent. For simplicity, we take the IR brane tension to be tuned and work in the flat slicing so that $z_\ir$ is constant in time.

In our setup with a detuned UV brane, 
the time evolution of $\phi$ comes from a potential $V(\phi)$ generated by the detuning of the UV brane tension. To calculate $V(\phi)$, we perform the trivial $z$ integration for UV brane localized term and get
\begin{align}
    V(\phi) &= \frac{L^4}{z_\uv^4} \delta T_\uv= \left( \phi + c \right)^2 \delta T_\uv\:,
\end{align}
where we have defined the constant
\begin{align}
    c = L^2/z_\ir^2
\end{align}
that parameterizes the IR brane. We note that the critical tension $T_c$ cancels against discontinuous terms coming from the bulk $R_5$, so only the detuning appears in the 4D potential. At this stage, we have the following form for the 4D effective action for $\phi$ coupled to gravity:
\begin{align}
    S_4 = \int d^4 x \sqrt{\left|g\right|} \left( \frac12 M_4^2\,\phi R_4 - \frac{w(\phi)}{\phi}(\partial\phi)^2 - V(\phi)\right)\: ,
    \label{eq:4D_action_radion_only}
\end{align}
where $w(\phi)$ is an undetermined function which we now fix by matching to the 5D solution. Note that the resulting action is that of a Brans-Dicke theory with the radion coupling to $R_4$.

We know the explicit time dependence of $z_\uv$ from the 5D solution, eq.~\eqref{eq:zuv5D}. From this we know the background solution $\phi = \phi_0(\tau)$ is
\begin{align}
    \phi_0(\tau) = \frac{L^2}{ (z_0 - \gamma \tau)^2} - c\:,
    \label{eq:phi_sol}
\end{align}
when the 4D metric is Minkowski.
Defining 
\begin{align}
    \eta = \tau - z_0/\gamma
    \label{eq:tau-eta-relation}
\end{align}
the background solution for $\phi$ becomes
\begin{align}
    \phi_0(\eta) = \frac{L^2}{ \gamma^2\eta^2} - c\:.
    \label{eq:phi-bkg-eta}
\end{align}
As discussed in sec.~\ref{subsec:5D-pic}, we choose $\gamma > 0$ so that $z_\uv(\tau) = z_0-\gamma \tau$ decreases with $\tau$, which makes $\eta < 0$. We will later identify $\eta$ with the conformal time in the 4D EFT. 
Requiring that the equations of motion are solved for $\phi = \phi_0(\tau)$ and $g_{\mu \nu} = \eta_{\mu \nu}$ fixes the function $w(\phi)$ to be
\begin{align}
    w(\phi) = - \frac{3 M_4^2 \phi }{4(\phi + c)} \,.
\end{align}
At this point we have derived the 4D EFT for the graviton $g_{\mu\nu}$ and radion $\phi$.  

It is convenient to transform the action in eq.~\eqref{eq:4D_action_radion_only} to the Einstein frame by rescaling the metric
\begin{align}
    g_{\mu \nu} \to g_{\mu \nu}/\phi \, ,
    \label{eq:WeylRescaling}
\end{align}
which leads to the action  
\begin{align}
    S =  \int d^4x \sqrt{-g} \left[ \frac12 M_4^2 \left(R_4  - \frac{3c}{2\phi^2(\phi + c)} (\partial \phi)^2 \right) 
	- \left( \frac{\phi + c}{\phi} \right)^2 \delta T_\uv \right]\: .
\end{align}
We can then define the canonically normalized radion, $\Pi(x)$, which is related to $\phi$ by:
\begin{align}
    &\Pi(x) = \sqrt{6} M_4 \arctanh \left(\sqrt{\frac{c}{c + \phi(x)}} \right) = \sqrt{6} M_4 \arctanh \left(z_\uv/z_\ir\right) \: , \:\:
    \phi(x) = c\,\csch^2 \left(\frac{\Pi(x)}{\sqrt{6} M_4} \right)\:.
    \label{eq:phiX}
\end{align}
Note that when the UV and IR branes are close, $z_\uv \sim z_\ir$, $\phi$ is small, and this corresponds to $\Pi \to \infty$. In the other extreme of $z_\ir \gg z_\uv$, $\Pi \to 0$. In terms of the canonically normalized radion $\Pi(x)$, the 4D action becomes
\begin{align}
    &S = \int d^4x \sqrt{-g} \left[ \frac12 M_4^2 R_4   - \frac12 (\partial \Pi)^2 
	- \cosh^4 \left(\frac{\Pi}{\sqrt{6} M_4}\right) \delta T_\uv\right]\:.
\end{align}
While this action was derived for a constant $\delta T_\uv$, it is expected to hold even when $\delta T_\uv$ is a function of other fields.

\section{UV-localized Inflaton}
\label{sec:UV-localized-inflaton}
We now consider an inflaton $\sigma$ localized on the UV brane, with a potential $v(\sigma)$. As $\sigma$ evolves, $v(\sigma)$ contributes to the tension of the UV brane. We take the bare tension on the UV brane to be the critical value $T_c$, so that $v(\sigma)$ is solely responsible for detuning and hence the UV brane motion.
The UV localized Lagrangian for $\sigma$ is given as
\begin{align}
    \mathcal{L}_\uv = -\frac{1}{2}(\partial \sigma)^2 - v(\sigma)\:.
    \label{eq:L-uv}
\end{align}
We assume $v(\sigma)$ takes a slow roll form so that initially $v(\sigma)$ is approximately constant.  We can therefore substitute $v(\sigma)$ for $\delta T_\uv$ into the expressions derived in section~\ref{sec:EFT-of-mismatched-branes}.

From the previous section, we already have part of the 4D EFT identified. To calculate the full EFT we need to calculate the contribution from $\sigma$. With the 5D metric given in eq.~\eqref{eq:5D_metric-general}, plugging $\mathcal{L}_\uv$ into eq.~\eqref{eq:5D_action-general} and doing the trivial $z$ integral, we get the following Lagrangian for $\sigma$: 
\begin{align}
	\mathcal{L}_\sigma = - \frac12 \frac{L^2}{z_\uv^2} g^{\mu \nu} \partial_\mu \sigma \partial_\nu \sigma
	- \frac{L^4}{z_\uv^4} v(\sigma)\, .
\end{align}
Recall that here $z_\uv$ is time-dependent and is related to $\phi$ and therefore $\Pi$ by eqs.~\eqref{eq:phidef},\eqref{eq:phiX}. 
In terms of $\Pi$ and $\sigma$, the 4D effective action is 
\begin{align}
    S &= \int d^4x \sqrt{-g} \left[ \frac12 M_4^2 R   - \frac12 (\partial \Pi)^2 
    -\frac12 f(\Pi) (\partial \sigma)^2
	- V(\Pi, \sigma) \right]\:, \\
    f(\Pi) &= \cosh^2 \left( \frac{\Pi}{\sqrt{6}M_4} \right) \:, \\
    V(\Pi, \sigma) &= \cosh^4\left( \frac{\Pi}{ \sqrt{6}M_4} \right) 
    v(\sigma)\:.
    \label{eq:full-action}
\end{align}
When the branes are close together ($\Pi \to \infty$), both the kinetic term and the potential for $\sigma$ are enhanced. In the opposite limit when the branes are far from each other ($\Pi \to 0$), the kinetic term for $\sigma$ has a canonical form and $V(\Pi,\sigma) \to v(\sigma)$. Note that we can't further redefine the fields $(\Pi, \sigma)$ to put them in a form with completely canonical kinetic terms, because the curvature in the field space is non-zero.

Of course brane inflation models already exist in the literature. Our model has some notable differences, two of which seem particularly relevant.  The form of the action we have derived (with  curvature in field space) has been studied in two-field inflation models for different choices of $f(\Pi)$ and $V(\Pi, \sigma)$ under the name of hyperbolic inflation~\cite{DiMarco:2002eb, DiMarco:2005nq, Lalak:2007vi, Tolley:2009fg, Brown:2017osf, Mizuno:2017idt, Garcia-Saenz:2018ifx, Garcia-Saenz:2019njm, Bjorkmo:2019aev, Fumagalli:2019noh}.  
Unlike these models, the form of  $f(\Pi)$ and $V(\Pi,\sigma)$ are  fixed in our scenario, making  our scenario more  predictive. Second, there is significant literature on utilizing the  motion of branes in a higher-dimensional space to generate an inflationary potential. In our setup, we have an inflaton field unrelated to the motion of branes: the brane motion is captured in the 4D theory by the radion $\Pi$, which is \textit{not} the inflaton.

\subsection{Equations of motion}

We next consider the equations of motion and the background solutions for the fields $\sigma, \Pi$ and the metric $g_{\mu\nu}$. In what follows, we denote the background solutions by $\sigma_0, \Pi_0, g_0^{\mu\nu}$. Given the action, the equations of motion for $\sigma$ and $\Pi$ are straightforward to obtain and are given as 
\begin{align}
    &\frac{1}{\sqrt{-g}}\partial_\mu \left( \sqrt{-g} g^{\mu \nu}  \partial_\nu \sigma \right)
    + \frac{1}{M_4}\sqrt{\frac{2}{3}} \tanh \left( \frac{\Pi}{\sqrt{6}M_4} \right) g^{\mu \nu}  \partial_\mu \Pi \partial_\nu \sigma
	- \cosh^2 \left( \frac{\Pi}{\sqrt{6}M_4} \right) \frac{\dd v( \sigma)}{\dd \sigma} = 0 \:,
    \label{eq:sigmaEOM}
    \\
    &\frac{1}{\sqrt{-g}}\partial_\mu \left( \sqrt{-g} g^{\mu \nu}  \partial_\nu \Pi \right)
	- \frac{1}{\sqrt{6} M_4} \sinh\left( \frac{\Pi}{\sqrt{6}M_4} \right)\cosh\left( \frac{\Pi}{\sqrt{6}M_4} \right) 
    \left( (\partial \sigma)^2 
    + 4 \cosh^2\left( \frac{\Pi}{\sqrt{6}M_4} \right) v(\sigma)\right) 
    =0 \, .
    \label{eq:XEOM}
\end{align}
In addition to these, we have the Einstein equations. We first consider the background solutions $\Pi_0, \sigma_0, g_0^{\mu\nu}$ which are time-dependent.
Assuming a flat Friedmann–Lemaître–Robertson–Walker (FLRW) background metric,
\begin{align}
    \dd s^2 = a^2(\eta)\left(-\dd \eta^2  +  \dd \vec{x}^2\right)\:,
\end{align}
the Einstein equations give 
\begin{align}
    \left(\frac{a'}{a}\right)^2 &= \frac{1}{3M_4^2}\left(\frac12 \Pi_0'^2 + \frac12 \cosh^2\left(\frac{\Pi_0}{\sqrt{6}M_4}\right)\sigma_0'^2 + a^2\,V(\Pi_0, \sigma_0)\right)\:, 
    \nonumber \\
    \frac{a''}{a}-2\left(\frac{a'}{a}\right)^2 
    &= -\frac{1}{2M_4^2} 
    \left(
    \Pi_0'^2 + \cosh^2\left(\frac{\Pi_0}{\sqrt{6}M_4}\right)\sigma_0'^2
    \right)\:,
    \label{eq:EE-bkg}
\end{align}
while the fields $\Pi_0, \sigma_0$ satisfy
\begin{align}
    &\sigma_0'' + 2\frac{a'}{a}\sigma_0'
    +\frac{1}{M_4}\sqrt{\frac23}\tanh\left(\frac{\Pi_0}{\sqrt{6}M_4}\right)\Pi_0'\sigma_0'
    +a^2 \cosh^2\left(\frac{\Pi_0}{\sqrt{6}M_4}\right)\partial_\sigma v(\sigma_0) = 0\:,
    \label{eq:bkg-eqn-sigma}
    \\
    &\Pi_0'' + 2\frac{a'}{a}\Pi_0'
    +
    a^2\frac{1}{\sqrt{6} M_4} \sinh\left( \frac{\Pi_0}{\sqrt{6}M_4} \right)\cosh\left( \frac{\Pi_0}{\sqrt{6}M_4} \right) 
    \left( \frac{\sigma_0'^2}{a^2} 
    + 4 \cosh^2\left( \frac{\Pi_0}{\sqrt{6}M_4} \right) v(\sigma_0)\right) 
    =0\:.
    \label{eq:bkg-eqn-pi}
\end{align}
From the scale factor $a$, the 4D Hubble $H$ is defined in the usual way:
\begin{align}
    H \equiv \frac{\dot{a}}{a} = \frac{a'}{a^2}\:,
\end{align}
and the equation of state is given as
\begin{align}
    w = \frac{2\,\Pi_0'^2 + 2\cosh^2\left( \frac{\Pi_0}{\sqrt{6}M_4} \right) \sigma_0'^2}{\Pi_0'^2 + \cosh^2\left( \frac{\Pi_0}{\sqrt{6}M_4} \right) \sigma_0'^2 + 2\,a^2\,V(\Pi_0,\sigma_0)} -1\:. 
\end{align}
We see that if the kinetic term of the radion dominates, $w\to 1$, leading to a period of kination, while if the inflaton dominates, $w\to -1$ and we have usual inflation. We will have more to say about this later. 

We now solve the coupled set of equations~\eqref{eq:bkg-eqn-sigma}, \eqref{eq:bkg-eqn-pi}, first in the limit of frozen inflaton and then in the limit of slowly rolling inflaton.

\subsection{Background solutions for frozen inflaton}

\label{subsec:bg_frozen}

We first consider $\sigma$ being frozen at a constant value, i.e. $\sigma_0$ being a constant. Instead of solving the equations directly, we note that in this limit we already know the background solutions. From the Weyl rescaling in eq.~\eqref{eq:WeylRescaling} and using eq.~\eqref{eq:phi-bkg-eta}, the background 4D metric is given as
\begin{align}
    g_0^{\mu\nu} &= \phi_0(\eta)\eta^{\mu\nu} \equiv a^2(\eta)\eta^{\mu\nu}\:.
    \label{eq:bkg-g-mu-nu}
\end{align}
Here the conformal scale factor $a(\eta)$ is given by
\begin{align}
    a^2(\eta) \equiv \phi_0(\eta) = \frac{L^2}{\gamma^2 \eta^2} - c\: \equiv \frac{1}{\Hsigma^2 \eta^2 } - c\,,
    \label{eq:scaleFactor}
\end{align}
and in the last equality, the parameter $\Hsigma$ is defined as:
\begin{align}
    \Hsigma^2 &= \frac{\gamma^2}{L^2} = \frac{v(\sigma_0)}{3 M_4^2}\:.
    \label{eq:Hf-vSigma}
\end{align}
Above, we have used $\delta T = v(\sigma_0)$ in eq.~\eqref{eq:zuv5D} and the relation between $M_4$ and $M_5$ from eq.~\eqref{eq:phidef}. $\Hsigma$ is purely sourced by the inflaton potential, and in the present discussion, since $\sigma_0$ is a constant, so is $\Hsigma$.

\begin{figure}[h]
\centering
\includegraphics[scale=0.55]{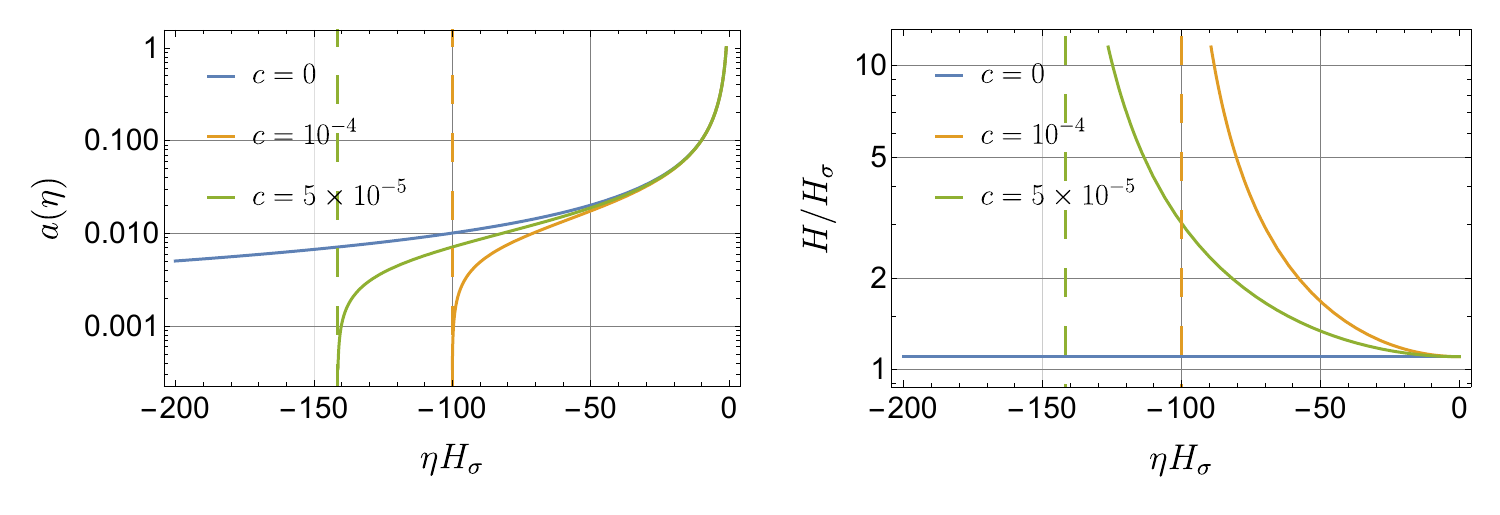}
\caption{\small{Left: Scale factor for $c = 0$ (blue), $10^{-4}$ (orange) and $5\times 10^{-5}$ (green). Dotted lines show where the scale factor goes to zero, corresponding to the UV and IR branes touching. Right: The effective 4D Hubble for $c = 0$ (blue), $10^{-4}$ (orange) and $5\times 10^{-5}$ (green). Dotted lines again correspond to the UV and IR branes touching. At late times, the scale factor looks like pure dS, but deviates at early times, and approaches zero at a finite value of $\eta$ that depends on $c$.}}
\label{fig:scaleFactor}
\end{figure}

Assuming constant $\Hsigma$, the 4D Hubble scale $H$, at a general conformal time $\eta$, can be derived from the expression for the scale factor:
\begin{align}
    H = \Hsigma\left(1-c \eta^2 \Hsigma^2\right)^{-3/2}\:.
    \label{eq:hubble-zero-slow-roll}
\end{align}
Fig.~\ref{fig:scaleFactor} shows the scale factor and $H/\Hsigma$ for a few values of $c$. We note that at late times ($\eta \to 0$) $H$ approaches $\Hsigma$. Crucially, we observe that $H$ grows at early times, and would be infinite at
\begin{align}
     \eta_{a \to 0} = -\frac{1}{\Hsigma\sqrt{c}}\:.
\end{align}
The presence of the IR brane, as parameterized by a non-zero $c$, therefore already makes an important difference: in contrast to typical inflationary models, $\eta$ can't be extrapolated to arbitrarily large negative values, but is bounded from below by $\eta_{a \to 0}$. This special value of $\eta$ corresponds in the 5D picture to the UV and the IR brane being on top of each other and the volume becoming zero. Clearly, this is where the 5D EFT can not be trusted. While an explicit UV model can answer what happens in this limit, here we will restrict ourselves to be away from  $\eta_{a \to 0}$. 
For a sensible 4D EFT we need $H$ to be smaller than the Planck scale $M_4$, which requires 
\begin{align}
    \eta  > \eta_* =  -\frac{1}{\Hsigma\sqrt{c}} \left(1-\left(\frac{\Hsigma}{M_4}\right)^{2/3}\right)^{1/2}\:.
    \label{eq:EFT-condition-on-eta}
\end{align}
If $\Hsigma/M_4 \ll 1$, $\eta_*$ is very close to but larger than $\eta_{a \to 0}$. In a given UV completion, the relevant cutoff $\Lambda$ may be at a scale below $M_4$, for example the string scale or the Planck scale in a higher-dimensional theory. In these cases, $M_4$ should be replaced by the appropriate scale $\Lambda$ in eq.~\eqref{eq:EFT-condition-on-eta}, and $|\eta_*|$ will be smaller.

In terms of $\Hsigma$ and $\eta$, the background solution $\Pi_0$ can be obtained using eq.~\eqref{eq:phiX}, and is given as
\begin{align}
    \Pi_0(\eta) = \sqrt{6}M_4 \arctanh\left(-\sqrt{c} \Hsigma \eta\right) \:.
    \label{eq:pi-bkg-sol}
\end{align}
As a consistency check, setting $\sigma_0'=0$, eq.~\eqref{eq:scaleFactor} and~\eqref{eq:pi-bkg-sol} satisfy eqns.~\eqref{eq:EE-bkg} and~\eqref{eq:bkg-eqn-pi}. Further, the relation~\eqref{eq:Hf-vSigma} is exactly the 4D Freedman equation $H^2 =\rho/3M_4^2$, which allows us to identify $\Hsigma$ with the Hubble observed by an observer on the UV brane. Note that for observers localized at different places in the bulk, the observed Hubble constant is different.\footnote{For example, for an observer localized on the IR brane, the induced metric is Minkowski. In the 4D EFT, IR localized fields have functions of $\Pi$ multiplying their kinetic terms, and the time dependence of $\Pi$ modifies their equation of motion. This ultimately reproduces the Minkowski space equations of motion, even though the 4D metric is approximately de Sitter.}

In the limit of $\sigma$ being frozen, the equation of state can be calculated explicitly, given $V(\Pi, \sigma)$ and the background solution for $\Pi$ in eq.~\eqref{eq:pi-bkg-sol}. It is given as
\begin{align}
    w = 2 c \,\Hsigma^2\eta^2 -1\:.
\end{align}
We see that at very early times (i.e. $c \Hsigma^2\eta^2 \to 1$ from below), the equation of state $w \to 1$, and the Universe experiences a kination phase driven by the kinetic energy of the fast rolling $\Pi$. At late times ($c \Hsigma^2\eta^2 \to 0$), the equation of state $w \to -1$ as the effect of $\Pi$ dilutes and the Universe undergoes standard slow roll inflation driven by $\sigma$. The condition for a decreasing co-moving horizon, $\partial_t(1/aH) < 0$ (or equivalently $w < -1/3$) translates to 
\begin{align}
    c \,\Hsigma^2\eta^2 < \frac13\:.
\end{align}
Note that the above condition gets corrected once we allow $\sigma_0$ to be time-dependent. 
Therefore, as we choose to restrict to the inflationary period, we consider only $c \,\Hsigma^2\eta^2 < 1/3$. It would also be interesting to allow $c \,\Hsigma^2\eta^2 > 1/3$. We leave this for future work.

\begin{figure}[h]
\centering
\includegraphics[scale=0.6]{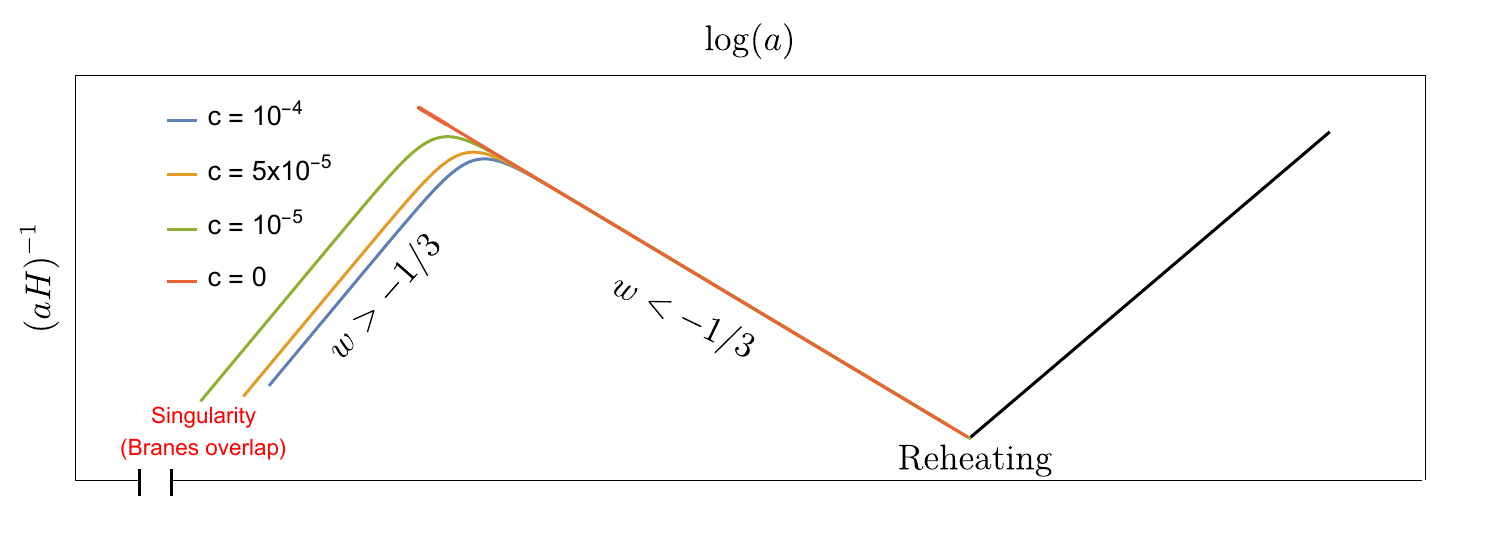}
\caption{\small{Comoving Hubble radius as a function of $\log(a)$. Compared to the $c=0$ case (red line), the behavior of comoving Hubble radius before reheating is different: inflation  occurs only for a finite period. This period is longer if $c$ is made smaller. For any $c\neq0$ the comoving Hubble radius and the scale factor go to zero at finite conformal time, corresponding to the branes overlapping in the 5D picture.
}}
\label{fig:comovingHubbleRadius}
\end{figure}

Figure~\ref{fig:comovingHubbleRadius} shows the comoving Hubble radius as a function of scale factor, for some values of $c$. For a non-zero $c$, the comoving Hubble radius is initially increasing and eventually starts to decrease, during inflation. There is a turn around corresponding to $w = -1/3$, which is pushed to an earlier time for smaller $c$. An important consequence of the turn around in $1/(aH)$ is that progressively smaller $k$ modes experience less e-folds of inflation, and some $k$ modes never experience any inflation.

\subsection{Slow roll solutions}
We now allow 
$\sigma$ to slowly vary as a function of time as in standard slow-roll inflation, so that $\Hsigma$ is time-dependent too. We define the slow-roll parameters
\begin{align}
    \epsilon_V \equiv \frac12M_4^2\left(\frac{\partial_\sigma v(\sigma)}{v(\sigma)}\right)^2\:,
    \:\:\eta_V \equiv M_4^2 \frac{\partial_\sigma^2\,v(\sigma)}{v(\sigma)}\:,\:\:
    \varepsilon \equiv -\frac{\partial_t \Hsigma}{\Hsigma^2} 
    = - \frac{\Hsigma'}{a\,\Hsigma^2}\:,
\end{align}
and work in the limit of these quantities being small in magnitude.

The expressions for $a$ and $\Pi$ remain unchanged to leading order except for the substitution $\Hsigma \to \Hsigma(\sigma)$, where $\Hsigma(\sigma)$ is now time-dependent due to the slowly varying $\sigma$. The parameters $\varepsilon$ and $\epsilon_V$ are related to each other by the equations of motion. For now we keep both $\varepsilon$ and $\epsilon_V$, and present the relation between them at the end of the section.
Differentiating the expression for the scale factor (eq.~\eqref{eq:scaleFactor}) and including the change of $\Hsigma$ parameterized by $\varepsilon$, the 4D Hubble gets corrected from eq.~\eqref{eq:hubble-zero-slow-roll} to
\begin{align}
    H &= \Hsigma\frac{1 + \eta  \Hsigma'/\Hsigma}{\left(1-c\,\eta^2\Hsigma^2\right)^{3/2}} 
    = \Hsigma\frac{1+\varepsilon\sqrt{1-c\,\eta^2 \Hsigma^2}}{\left(1-c\,\eta^2 \Hsigma^2\right)^{3/2}}\:.
    \label{eq:Hubble-slow-roll-corrected}
\end{align}

To obtain the equation of motion for $\sigma_0$ in the slow roll limit, we can substitute the zeroth order background solution for $\Pi_0$ and $g_0^{\mu\nu}$ (see eqns.~\eqref{eq:pi-bkg-sol}, \eqref{eq:bkg-g-mu-nu}) in eq.~\eqref{eq:bkg-eqn-sigma} to obtain
\begin{align}
    \ddot{\sigma}_0+ \dot{\sigma}_0\left(\frac{\dot{a}}{a}-\frac{2}{\eta\,a}\right) + \frac{\partial_\sigma v(\sigma_0)}{\Hsigma^2\,\eta^2\,a^2} = 0\:.
    \label{eq:sigma-eom-tCoordinate}
\end{align}
Note that the equation depends on $c$ through $a(\eta)$ (see eq.~\eqref{eq:scaleFactor}). However, expressions take a simpler form in the $\eta$ coordinate, in which the equation satisfied by $\sigma$ becomes
\begin{align}
    \sigma_0'' 
    - \frac{2}{\eta}  \sigma_0' 
	+ \frac{1}{\Hsigma^2 \eta^2} \partial_\sigma v( \sigma_0 )
    &= 0 \:.
    \label{eq:sigma-eom-etaCoordinate}
\end{align}
Note that $c$ has dropped out of the equation, and this is the equation of motion for a scalar field in pure de Sitter space. 
This is expected since a field on the UV brane should see the induced metric on the brane, which on the UV brane is exactly de Sitter.

We now consider the rate of change of the background solutions, which allows us to identify when the corresponding fields are rolling slowly or quickly.
To obtain the rate of change of $\sigma$, we use eq~\eqref{eq:sigma-eom-tCoordinate} and drop $\ddot{\sigma}_0$ in the slow roll limit to obtain
\begin{align}
    \dot{\sigma}_0 = -\frac{\partial_\sigma v(\sigma)}{\Hsigma^2\,\eta^2\,a^2}\left(\frac{\dot{a}}{a}-\frac{2}{\eta\,a}\right)^{-1}\:.
\end{align}
Substituting the explicit form of $a$ and using the definition of $\epsilon_V$, we get
\begin{align}
    \frac{\dot{\sigma}_0}{M_4 H} &= -\sqrt{2\epsilon_V}\,\bigg(1- c\,\eta^2 \Hsigma^2\bigg)^2\left(1-\frac23 c\,\eta^2 \Hsigma^2\right)^{-1} \:. 
    \label{eq:sigma-dot}
\end{align}
The rate is suppressed by the slow roll parameter $\epsilon_V$ as expected.

For the radion, given the explicit solution in eq.~\eqref{eq:pi-bkg-sol}, it is straightforward to obtain
\begin{align}
    \frac{\dot{\Pi_0}}{ M_4 H} = 
    \frac{\sqrt{6c}  \, \eta \Hsigma\left(1+\varepsilon\sqrt{1-c\,\eta^2 \Hsigma^2}\right)}{1-\varepsilon\sqrt{1-c\,\eta^2 \Hsigma^2}}
    = \sqrt{6c}\,\eta \Hsigma\left(1+2\varepsilon\sqrt{1-c\eta^2 \Hsigma^2}\right) + \cdots
    \:.
    \label{eq:Pi-dot}
\end{align}
We see that $\dot{\Pi_0}$ is proportional to $\eta$, so when the branes are close ($c\,\eta^2 \Hsigma^2 \to 1$) the evolution is rapid whereas evolution is slow at late times ($\eta \to 0$). The latter time dependence includes a term proportional to both $\sqrt{c}$ and $\varepsilon$ which is doubly suppressed at late times. The overall factor of $\sqrt{c}$ in eq.~\eqref{eq:Pi-dot} can be understood as follows: in the limit of $c\to0$ (i.e. the IR brane sent to infinity in the 5D picture), the radion $\Pi$ decouples from the theory and 
both $\Pi$ and $\dot{\Pi}$ vanish (see eq.~\eqref{eq:phiX}). 

To obtain the relation between $\varepsilon$ and $\epsilon_V, \eta_V$, we consider the Einstein equation~\eqref{eq:EE-bkg}, written in $t$ coordinate:
\begin{align}
    \dot{H} = -\frac{1}{2 M_4^2}
    \left(\dot{\Pi}^2 + \cosh^2\left(\frac{\Pi}{\sqrt{6} M_4}\right)\dot{\sigma}^2\right)\:.
\end{align}
Expanding 
$H$ to linear order in the slow roll parameters we find
\begin{align}
    \varepsilon =  \epsilon_V 
    \frac{\left(1-c\, \Hsigma^2 \eta^2\right)^{5/2}}
    {\left(1-\frac23 \,c\, \Hsigma^2 \eta^2\right)^2 \left(1+ 4 \,c\, \Hsigma^2 \eta^2\right) }
    \approx
    \epsilon_V\left(1-\frac16 c\Hsigma^2\eta^2\right)
    \:.
    \label{eq:epsilon-vs-epsilon}
\end{align}
At late times, $c\,\Hsigma^2\eta^2 \to 0$, the dynamics approaches the expectations for a single field model and we recover the standard result $\varepsilon = \epsilon_V$.

\subsection{The start and end of inflation}
The period of cosmology that can be studied in the EFT
is restricted to lie between
\begin{align}
    \eta_* \leq \eta \leq \eta_\text{end}\:, 
    \label{eq:allowed-period-of-cosmology}
\end{align}
where $\eta_*$ is defined in eq.~\eqref{eq:EFT-condition-on-eta} and $\eta_\text{end}$ is when inflation ends (where $|\Hsigma\eta_\text{end}| \ll 1$ to generate sufficient number of e-folds). The lower bound in eq.~\eqref{eq:allowed-period-of-cosmology} comes from requiring that the 4D Hubble scale is smaller than the Planck scale. As we saw in sec.~\ref{subsec:bg_frozen}, the equation of state is $\eta$ dependent. An era of inflation requires $w < -1/3$, which translates to
\begin{align}
    \eta > -\frac{1}{\Hsigma\sqrt{3c}}\:.
\end{align} 
This translates to the initial condition in the 5D picture to be such that the two branes start a finite distance apart, with $z_\uv/z_\ir < 1/\sqrt{3}$ (see eq.~\eqref{eq:phiX}). This condition also ensures that the field $\Pi$ does not take Planckian values.
This means that the period of \textit{inflationary} cosmology that can be studied in our setup is during 
\begin{align}
    -\frac{1}{\Hsigma\sqrt{3c}} \leq \eta \leq \eta_\text{end}\:.
    \label{eq:allowed-period-of-inflation}
\end{align}

Inflation is driven by the detuned UV brane tension (meaning away from the value that would give flat space), which is set by the potential of the UV localized inflaton. In the slow-roll region when the inflaton  moves slowly in the flat part of its potential, the UV brane detuning is approximately constant and the UV brane moves with a nearly constant rate (w.r.t. time $\tau$ or $\eta$) in the 5D bulk. Towards the end of inflation the inflaton rolls to its minimum and UV tension relaxes to the tuned value (which is the 5D condition corresponding to the inflaton rolling to flat space), at which point the UV brane stops moving in the bulk and inflation ends. Reheating then occurs as the inflationary energy is transferred to radiation and matter fields. In this paper we will assume instantaneous reheating.

In the 5D picture, inflation ends with the inflaton oscillation and reheating, but the radion can in principle still be moving. Whether this is true depends on the radion stabilization scale~\cite{Lust:2025vyz}. One possibility is that the radion is heavy, in which case it would be stabilized well before inflation ends and the the scenario is equivalent to 4D inflation. However, if the radion mass is less than $H$ at the start of inflation but greater than $H_\sigma$ it would lead to the same sort of deviations at early times that we derive below. If the stabilized radion mass is lighter than $H_\sigma$, which is perhaps less tuned than the previous option, the inflationary predictions are those derived below so long as $v(\sigma)$ is bigger than the stabilizing potential, in which case inflation happens at energies much higher than the IR scale corresponding to the cutoff of the low energy 4D EFT. 

\begin{figure}[h]
\centering
\includegraphics[scale=0.107]{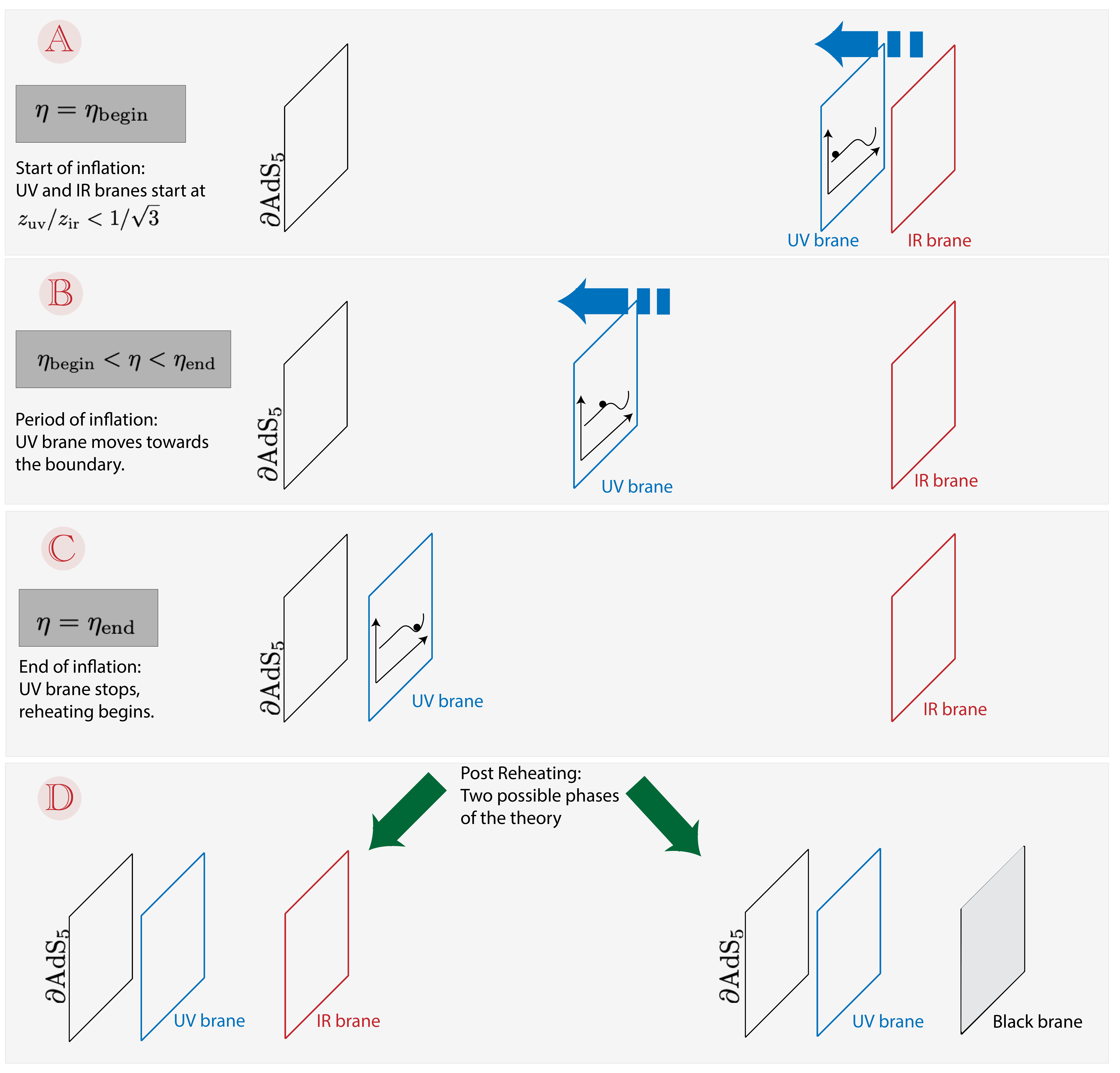}
\caption{\small{Top three panels show three snapshots of the UV brane motion in Minkowski slicing, correlated with the motion of the localized inflaton: (A) the beginning of inflation when the UV and IR brane are close to each other and UV brane begins to move, (B) continuation of the slow roll period with UV brane moving toward the boundary, and (C) the end of inflation when the inflaton settles to its minimum, the UV brane tension goes to the critical value and the UV brane comes to rest. Post reheating (D), the theory can end up in two different phases: with an IR brane or with a black brane in the IR. The IR brane may or may not move in accordance with the location of the minimum of the stabilizing potential, and the black brane phase would further evolve to the stabilized phase.}}
\label{fig:5Dpicture}
\end{figure}

Depending on the reheat temperature and dynamics, the theory post inflation can end up in two different phases: the confined phase, or the deconfined phase. For reheat temperatures below the radion mass, we expect to be in the stabilized confined phase. The confined phase is dual to the 5D geometry with a static UV and IR brane, and the location of the IR brane sets the confinement scale relative to the UV scale. With higher reheat temperature, as might be expected, the resulting cosmology depends on whether the IR degrees of freedom are reheated to this temperature.  If they are,  the theory can enter a black brane phase, which is dual to a 5D geometry with a UV brane and a black brane in the IR, and with the location of the black brane setting the temperature of the dual theory~\cite{Creminelli:2001th}.  In this case, the theory enters the confined phase only after a phase transition, the dynamics of which depends on further details of the theory~\cite{Creminelli:2001th, Kaplan:2006yi, Hassanain:2007js, Konstandin:2010cd, Dillon:2017ctw, Bunk:2017fic, vonHarling:2017yew, Baratella:2018pxi, Fujikura:2019oyi, Agashe:2020lfz, Csaki:2023pwy,Eroncel:2023uqf, Mishra:2023kiu, Mishra:2024ehr,Gherghetta:2025krk}. 
However, it is perhaps more natural for the inflaton to preferentially reheat only the Standard Model states, which are on the same brane as the inflaton in our scenario. In this case, the KK modes are  heated  to only a very low temperature and the brane locations will be set by the stabilizing potential as above.\footnote{Another possibility that avoids a confining phase transition is if the temperature dependence of the radion potential means that the RS phase remains metastable after reheating~\cite{Agrawal:2021alq}.}
A cartoon showing various snapshots of this dynamics is shown in fig.~\ref{fig:5Dpicture}.

\section{Linear Fluctuations: Scalar Modes}

\label{sec:linear-fluctuations}

We now consider scalar fluctuations around the background solutions derived in section~\ref{sec:UV-localized-inflaton}. We identify the adiabatic and isocurvature fluctuations and solve the coupled system of equations numerically. The main new features of our setup are a modified scale factor due to the correction from the presence of the IR brane and a fixed starting point for inflation.
We also provide an analytical approximation for the solutions in certain limits, adapting the procedure detailed in ref.~\cite{Lalak:2007vi} which also considered a two-field model with negative field-space curvature.
Throughout this section, we present expressions to linear order in the slow-roll parameters $\epsilon_V, \eta_V$. In sec.~\ref{sec:adiabatic-and-cmb} we calculate the power spectrum numerically and present analytical fits to the numerical result. We also quantify deviations from the single-field case (corresponding to setting $c = 0$) for the CMB angular modes.

There are three sources of scalar fluctuations, two from the fields $\sigma$ and $\Pi$ and one from the metric itself. However only two combinations are physical. Working in longitudinal gauge and in the absence of off-diagonal spatial components of the stress-energy tensor, as is the case for scalar matter, the metric including the scalar fluctuation $\Phi$ is given as
\begin{align}
    \dd s^2 = -(1+2\Phi)\dd t^2 + a^2(t)(1-2\Phi)\dd x^2\:, 
\end{align}
while the fluctuations $\delta\Pi, \delta\sigma$ are defined by expanding the fields around the background solution:
\begin{align}
    \Pi(x, t) &= \Pi_0(t) + \delta\Pi(x, t)\:,\\
    \sigma(x, t) &= \sigma_0(t) + \delta\sigma(x,t)\:.
\end{align}
Under a time reparameterization $t \to t'(t)$, the quantities $\delta\Pi, \delta\sigma$ and $\Phi$ change. There are two combinations that are invariant under these reparameterizations, the so-called Mukhanov-Sasaki variables: 
\begin{align}
    Q_\Pi(x,t) &= \delta\Pi + \frac{\dot{\Pi}}{H}\Phi\:,\\
    Q_\sigma(x,t) &= \delta \sigma + \frac{\dot{\sigma}}{H} \Phi\:,
\end{align}
where dots here refer to $t$ derivatives. Since the background respects 3D translational symmetry, we work with the 3D Fourier mode functions $Q_{\Pi, \textbf{k}}, Q_{\sigma, \textbf{k}}$ (with mass dimension $-1/2$),  
and suppress the subscript $\textbf{k}$ in what follows for notational simplicity.

\subsection{Equations of motion}
\label{subsec:EOM-Q-pi-Q-sigma}
The equations of motion for $Q_\Pi, Q_\sigma$ are in general mixed, and have the form
\begin{align}
    \ddot Q_\Pi + \left(3H+B_{\Pi\Pi}\right) \dot{Q_\Pi} + B_{\Pi\sigma}\dot Q_\sigma + \left(\frac{k^2}{a^2} + C_{\Pi\Pi}\right)Q_\Pi + C_{\Pi\sigma} Q_\sigma &= 0\:, 
    \nonumber
    \\
    \ddot Q_\sigma + \left(3H+B_{\sigma\sigma}\right) \dot{Q_\sigma} + B_{\sigma\Pi}\dot Q_\Pi + \left(\frac{k^2}{a^2} + C_{\sigma\sigma}\right)Q_\sigma + C_{\sigma\Pi} Q_\Pi &= 0\:,
    \label{eq:EOM-FluctuationsAlongFields}
\end{align}
where the expression for the constants $B_{ij}, C_{ij}$ for $i, j$ in $(\Pi, \sigma)$, understood to be evaluated on the background value of $\Pi, \sigma$, are given in the appendix~\ref{app:EOM-fluctuations-along-field-direction}. The quantities $C_{\Pi\Pi}, C_{\sigma\sigma}$ are the mass-squared terms for the respective fluctuations. There is also a mass mixing from $C_{\Pi\sigma}, C_{\sigma\Pi}$, and kinetic mixing from $B_{\Pi\sigma}, B_{\sigma\Pi}$ terms.
In terms of $\epsilon_V$ and $c$, the full expressions for the mass and kinetic mixings are given in appendix~\ref{app:EOM-fluctuations-along-field-direction}. 

The early time behavior of these quantities comes from $H$ becoming large (see eq.~\eqref{eq:hubble-zero-slow-roll}). At late times, $c \Hsigma^2 \eta^2 \to 0$, we find that the model reduces to that of a single-field model, which can be seen as taking $c = 0$ sends the IR brane to infinity and decouples the radion. We now observe a couple of things for the terms in eq.~\eqref{eq:EOM-FluctuationsAlongFields}, focusing on the full range $0 \leq c \Hsigma^2 \eta^2 \leq 1$.

\begin{itemize}
    \item $C_{\Pi\Pi}/H^2$ is small at early times and approaches a positive value towards the end. The leading behavior of $C_{\Pi\Pi}/H^2$ can be understood by first setting $\epsilon_V = 0$, which gives
    \begin{align}
    \frac{C_{\Pi\Pi}}{H^2}
    &= 
    2\left(1-c\, \Hsigma^2 \eta^2\right)\:.
    \label{eq:mass-cPiPi}
    \end{align}
    We see clearly that at early times $C_{\Pi\Pi}/H^2$ approaches zero and at late time approaches 2, which comes from the $\partial_\Pi^2 V(\Pi, \sigma)$ term (all the other terms are subleading). $C_{\Pi\Pi}/H^2$ = 2 is the expected result for a conformally coupled scalar. Using the parameterization in eq.~(5.19) of ref.~\cite{Karch:2020iit} and its further identification with the action in ref.~\cite{Chacko:2013dra}, one sees that the action indeed has the form of a conformally coupled scalar with coupling $(1/12)R_4\varphi$ for a canonically normalized scalar $\varphi$. 
    
    To leading order in the slow roll parameters, the early and late time limits are given as
    \begin{align}
    \frac{C_{\Pi\Pi}}{H^2}
    &\underset{c\, \Hsigma^2 \eta^2\to 1, \epsilon_V  \ll 1}{=} 
    2(1-c\, \Hsigma^2 \eta^2) - \frac{516}{5}\epsilon_V(1- c\, \Hsigma^2 \eta^2)^3
    \\
    \frac{C_{\Pi\Pi}}{H^2}
    &\underset{c\, \Hsigma^2 \eta^2\to 0, \epsilon_V \ll 1}{=} 2-\frac{13}{3}\epsilon_V + \cdots \:,
    \end{align}
    which shows that $C_{\Pi\Pi}/H^2$ starts close to zero and saturates to a value slightly different from 2 at late times.
    
    \item $C_{\sigma\sigma}/H^2$ also starts close to zero at early times, and goes to a value much smaller than unity, being suppressed by slow-roll parameters. Expanding to linear order in the slow-roll parameters we get
    \begin{align}
        \frac{C_{\sigma\sigma}}{H^2}
        &= 3\left(1-c\, \Hsigma^2 \eta^2\right)^2
        \left(
         \eta_V
        -2\epsilon_V  \frac{\left(1-c\, \Hsigma^2 \eta^2\right) \left(1-\frac13 c\, \Hsigma^2 \eta^2\right)}{\left(1-\frac23 c\, \Hsigma^2 \eta^2\right)^2}
        \right)\:.
    \end{align}
    The overall factor of $(1-c\, \Hsigma^2 \eta^2)$ ensures $C_{\sigma\sigma}/H^2$ starts at zero. At late times, $c\, \Hsigma^2 \eta^2 \to 0$ and $C_{\sigma\sigma}/H^2 = 3(\eta_V-2\epsilon_V)$, which is the expected result in single-field models~\cite{Baumann:2022mni}.
    
    \item  The mass-mixing terms $C_{\Pi\sigma}/H^2, C_{\sigma\Pi}/H^2$ approach zero at both early and late times, and are suppressed by the slow-roll parameter $\epsilon_V$. This is seen by the respective expressions in app.~\ref{app:EOM-fluctuations-along-field-direction}, where each term in their expression goes to zero in the slow-roll limit, as well as in the $c\, \Hsigma^2 \eta^2\to0$ and $c\, \Hsigma^2 \eta^2\to1$ limits. 
    
    \item The kinetic mixing terms $B_{\Pi\sigma}/H^2, B_{\sigma\Pi}/H^2$ also approach zero at both early and late times, and are suppressed by the slow-roll parameter $\epsilon_V$. This is seen by the respective expressions in app.~\ref{app:EOM-fluctuations-along-field-direction}, where each contributing term goes to zero in the slow-roll limit, as well as in the $c\, \Hsigma^2 \eta^2\to0$ and $c\, \Hsigma^2 \eta^2\to1$ limits. 
    \item The diagonal term $B_{\sigma\sigma}$ is negative, starting at $-2H$ at early times and going to zero at late times. This term is compensated by the Hubble friction term $3H$ to give an overall positive friction term.
\end{itemize}

Given the mass matrix elements $C_{ij}$, we can diagonalize the mass matrix to get the two mass eigenvalues (ignoring the mixed Hubble friction terms) as a function of time. Since the mixing terms are suppressed by the slow roll parameters, the effect of mixing is small. We therefore have two scalar modes, with mass-squared approximately given by $C_{\Pi\Pi}$ and $C_{\sigma\sigma}$ respectively. See app.~\ref{app:EOM-fluctuations} for more details. 

\subsection{Adiabatic and entropy components}
\label{subsec:adiabatic-entropy-fluctuations}

\begin{figure}[h]
\centering
\includegraphics[scale=0.7]{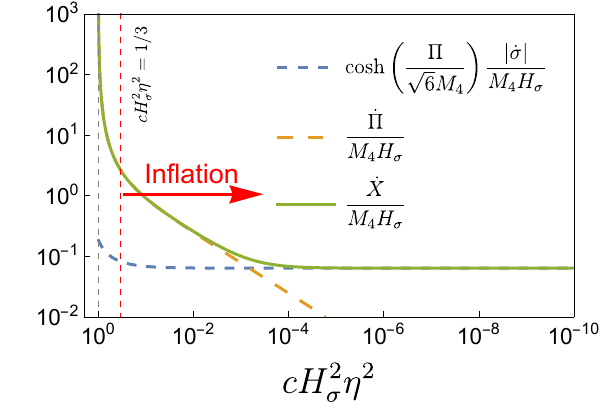}
\:\:
\includegraphics[scale=0.72]{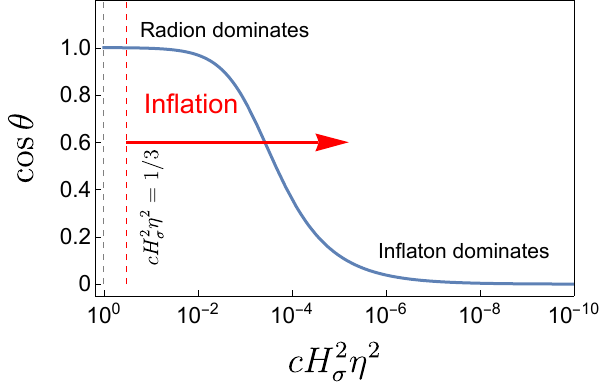}
\caption{\small{Left: $\dot{\Pi}$ (orange dashed) and $\cosh(\Pi/(\sqrt{6}M_4))\left|\dot{\sigma}\right|$ (blue dashed) in units of $M_4 \Hsigma$, as a function of $c\Hsigma^2\eta^2$. Also shown is the quantity $\dot{X}$ defined in eq.~\eqref{eq:small-theta}. At early times $\dot{\Pi}$ dominates while at late times $\cosh(\Pi/(\sqrt{6}M_4))\dot{\sigma}$ dominates. Right: $\cos \theta$ that parameterizes the rotation from the fluctuations of $\sigma, \Pi$ to the adiabatic and entropy components as a function of $c\Hsigma^2\eta^2$. At early times, $\cos \theta \to 1$ and the adiabatic fluctuations are mostly along $\Pi$, while at late times $\cos \theta \to 0$ and the adiabatic fluctuations are mostly along $\sigma$. Both plots indicate the region $c\Hsigma^2\eta^2>1/3$ when the spacetime inflates. In these plots we have taken $\epsilon_V = 0.002$.}}
\label{fig:theta-angle}
\end{figure}

To connect to late-time observables, it is useful to decompose the scalar perturbations along the direction parallel to the background trajectory and orthogonal to it~\cite{Gordon:2000hv, GrootNibbelink:2000vx, Bartolo:2001rt, GrootNibbelink:2001qt}. We denote these two directions as the $r$ (for curvature) and the $s$ (for entropy) directions respectively. We define
\begin{align}
    Q_r = \cos \theta \, Q_\Pi + \sin \theta \, \cosh(\frac{\Pi}{\sqrt{6}M_4}) \, Q_\sigma\:, \qquad 
    Q_s = -\sin \theta \, Q_\Pi + \cos \theta \, \cosh(\frac{\Pi}{\sqrt{6}M_4}) \, Q_\sigma\:,
    \label{eq:Q-pi-q-sigma-to-Q-r-Q-s}
\end{align}
where
\begin{align}
    \cos \theta = \frac{\dot{\Pi}}{\dot{X}}
    \:,\:\: \sin \theta = \cosh\left(\frac{\Pi}{\sqrt{6}M_4}\right)\,\frac{\dot{\sigma}}{\dot{X}}\:,\:\: \dot{X} = \sqrt{\dot{\Pi}^2 + \cosh^2\left(\frac{\Pi}{\sqrt{6}M_4}\right) \dot{\sigma}^2}\:.
    \label{eq:small-theta}
\end{align}
$Q_r$ denotes the gauge invariant fluctuation along the tangent to the background trajectory, and $Q_s$ denotes the gauge invariant fluctuation orthogonal to the tangent. Note that $\dot{X}$ is just defined for convenience, and is not a field or time derivative of a field. 
The factors of $\cosh(\Pi/(\sqrt{6} M_4))$ are due to the curvature in the $(\Pi, \sigma)$ field space, see discussion near eq.~\eqref{eq:full-action}.
At early times $\dot{\Pi}$ is large compared to $\cosh(\frac{\Pi}{\sqrt{6}M_4})\dot{\sigma}$, but is subleading at late times. This means that at early times, $\dot{X}\approx \dot{\Pi}$, while at late times $\dot{X} \approx \cosh(\frac{\Pi}{\sqrt{6}M_4})\dot{\sigma}$. Using eqns.~\eqref{eq:sigma-dot},\eqref{eq:Pi-dot}, for arbitrary $\eta$ we get
\begin{align}
    \left(\frac{\dot{X}}{M_4 H}\right)^2 =  \frac{6c\,\eta^2\Hsigma^2}{(1-c\,\eta^2 \Hsigma^2)^3} + \frac{2\epsilon_V(1-c\,\eta^2 \Hsigma^2)}{(1-\frac23 c\,\eta^2 \Hsigma^2)^2}\:.
    \label{eq:X}
\end{align}
The left panel of fig.~\ref{fig:theta-angle} shows field derivatives as a function of $c \Hsigma^2 \eta^2$, and $\dot{X}/(M_4 H)$ between its two limiting values.  The right panel of fig.~\ref{fig:theta-angle} shows $\cos \theta$ as a function of $c\Hsigma^2\eta^2$. We see that at early times $\cos \theta \to 1$, reflecting the convergence of $\dot{X}$ and $\dot{\Pi}$ while at late times $\cos \theta$  goes to $0$ reflecting constant $\Pi$. This is the expectation: at early times the classical trajectory is mostly along the radion, while at late times it is mostly along the inflaton.

\subsection{Mode functions and choice of quantization}
Starting with the equations of motion for the fluctuations $(Q_\Pi, Q_\sigma)$ in sec.~\ref{subsec:EOM-Q-pi-Q-sigma}, we can derive the coupled equations for the adiabatic mode function $Q_r$ and the isocurvature mode function $Q_s$ (using eq.~\eqref{eq:Q-pi-q-sigma-to-Q-r-Q-s}).  
We  solve the equations for $Q_r, Q_s$ first in the approximation that the time-dependent coefficients in the differential equation can be replaced by their value at Hubble crossing, and then compare this approximation to the solution obtained numerically without any approximations. In the process we will clarify the the choice of 
initial condition for the mode functions for $Q_r, Q_s$, and the choice of vacuum.

We work with the dimensionless quantities
\begin{align}
    x = \eta \Hsigma\:,\:\:\kappa =  k/\Hsigma\:, q_r = \Hsigma^{1/2}Q_r\:,\:\: q_s = \Hsigma^{1/2} Q_s\:.
\end{align}
Leaving the details to appendix~\ref{app:EOM-fluctuations}, the coupled equations satisfied by $q_r, q_s$ are given as
\begin{align}
    &q_r''+b_{rr} q_r' + b_{rs}q_s' + (\kappa^2 + c_{rr})q_r + c_{rs} q_s = 0\:,
    \nonumber \\
    &q_s''+b_{ss} q_s' + b_{sr}q_r' + (\kappa^2 + c_{ss})q_s + c_{sr} q_r = 0\:.
    \label{eq:qr-qs-eqn}
\end{align}
The quantities $b_{ij}, c_{ij}, i, j \in (r,s)$ are functions of $\eta, c, \epsilon_V, \eta_V$ and $x$. Explicit expressions for them are given in appendix~\ref{app:EOM-fluctuations-along-field-direction}. Once the theory is quantized and $q_r, q_s$ are promoted to operators $\hat{q}_r, \hat{q}_s$, we have
\begin{align}
    \hat{q}_r = q_r \hat{a}_{\bf{k}} + \text{ h. c.}\:,\:\: \hat{q}_s = q_s \hat{b}_{\bf{k}} + \text{ h. c.}\:.
\end{align}
Here, $\hat{a}_{\bf{k}}, \hat{b}_{\bf{k}}$ are annihilation operators that annihilate the vacuum $\left|\Omega\right>$,
\begin{align}
    &\hat{a}_{\bf{k}} \left |\Omega \right> = 0\:,\:\:\hat{b}_{\bf{k}} \left| \Omega \right> = 0\:,
\end{align}
and satisfy
\begin{align}
    &[\hat{a}^{\phantom{\dagger}}_{\bf{k}}, \hat{a}^\dagger_{{\bf{k'}}}] = [\hat{b}^{\phantom{\dagger}}_{\bf{k}}, \hat{b}^\dagger_{{\bf{k'}}}]= \delta^{(3)}(k-k')\:, \ \ [\hat{a}^{\phantom{\dagger}}_{\bf{k}}, \hat{b}^\dagger_{{\bf{k'}}}] = 0\:.
\end{align}

We choose the initial conditions for $q_r, q_s$ to be a plane wave at $x= x_i$:
\begin{align}
    q_r(x_i) = \frac{1}{a(x_i)\sqrt{2\kappa}}e^{-i \kappa x_i}\:,\:\:
    q_s(x_i) = \frac{1}{a(x_i)\sqrt{2\kappa}}e^{-i \kappa x_i}\: .
    \label{eq:qr-qs-ic}
\end{align}
In the usual case, the boundary condition is imposed asymptotically, i.e. for $x_i \to -\infty$.
The justification for this choice comes from the fact that for large $k$ values, eqns.~\eqref{eq:qr-qs-eqn} become degenerate and $q_r, q_s$ decouple. The solutions for $a(x) q_r(x), a(x) q_s(x)$ are proportional to plane wave solutions $e^{\pm i k \eta}$ and we choose the positive energy solution $e^{-ik\eta}$ like in Minkowski space. The physical intuition is that for large $k$ (short length scales), the effect of Hubble is negligible, and the vacuum locally looks like the Minkowski vacuum. Large $k$ and large $|\eta|$ are interchangeable because they appear in the product, and correspond to the limit $x_i \to -\infty$. Since there is no restriction in going to larger $|\eta|$ in the usual case, any $k$ mode can be decoupled if one goes early enough in time.

Compared to the usual cases where inflation does not have a beginning, there are important differences here that we point out now. First, note that in our setup inflation  begins only after $x = \eta \Hsigma =  -1/\sqrt{3c}$, so $x_i$ should not be taken to be smaller than $-1/\sqrt{3c}$ in eq~\eqref{eq:qr-qs-ic}. Therefore, if we want to restrict to an inflationary period, we should choose an initial value $x_i$ such that
\begin{align}
   x_i \geq -1/\sqrt{3c} \: .
\end{align}
In this work we will choose a few values of $x_i \geq -1/\sqrt{3c}$ and quantify the effect of different choice of $x_i$ on the observables of interest.
Furthermore, since the solution is singular at $x = -1/\sqrt{c}$ (the scale factor goes to zero, corresponding to the branes being on top of each other in the 5D picture), we are bounded from below in the choice of $x_i$. The lower bound on $x_i$ means that for small enough $k$, the solution for $a(x)q_r(x)$ and $a(x)q_s(x)$ are not just plane waves. In the calculations that follow, we will choose to work with the initial conditions~\eqref{eq:qr-qs-ic} for \textit{all} $k$. This choice is well justified for large $k$, since in that limit we know it corresponds to the Bunch-Davies state that minimizes the energy. For smaller $k$ values, it corresponds to an excited state. This is further discussed in section~\ref{subsec:initial-conditions}. 
The coefficients $b_{ij}, c_{ij}$ appearing in eq.~\eqref{eq:qr-qs-eqn} are in general $x$ dependent, and require the equations to be solved numerically. The equations can in principle be solved in the approximation that for a given $\kappa$, the coefficients $b_{ij}, c_{ij}$ are replaced by their value at Hubble crossing~\cite{Lalak:2007vi}. 
Since we expect to recover the single field result for large $k$ (which has an analytic form), using this approximation (which also allows an analytic form) provides a useful check on our results in the appropriate limit. However, we note that 
while it is useful to have an analytical expression to understand the parametric dependence of our result, this approximation is good in our model only for large $k$, and fails as we go to smaller $k$ values.  In the next subsection we will use this approximation to derive the analytical solutions for $q_r, q_s$ and compare them to the numerical solutions. 

\subsection{Analytical solution}
\begin{figure}[h]
\centering
\includegraphics[scale=0.53]{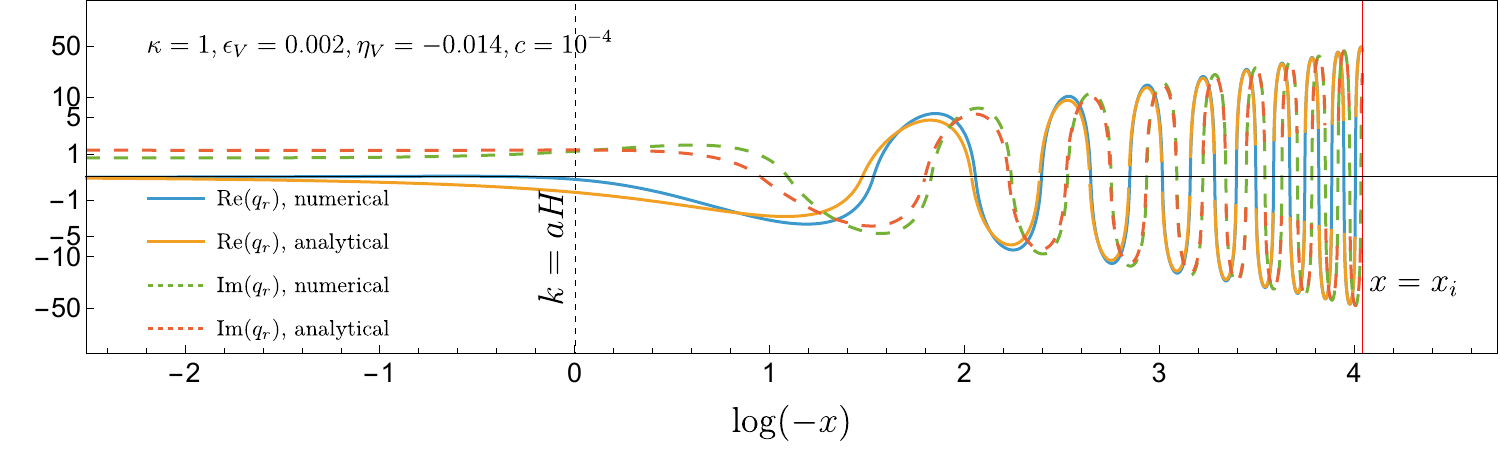}
\:\:
\includegraphics[scale=0.53]{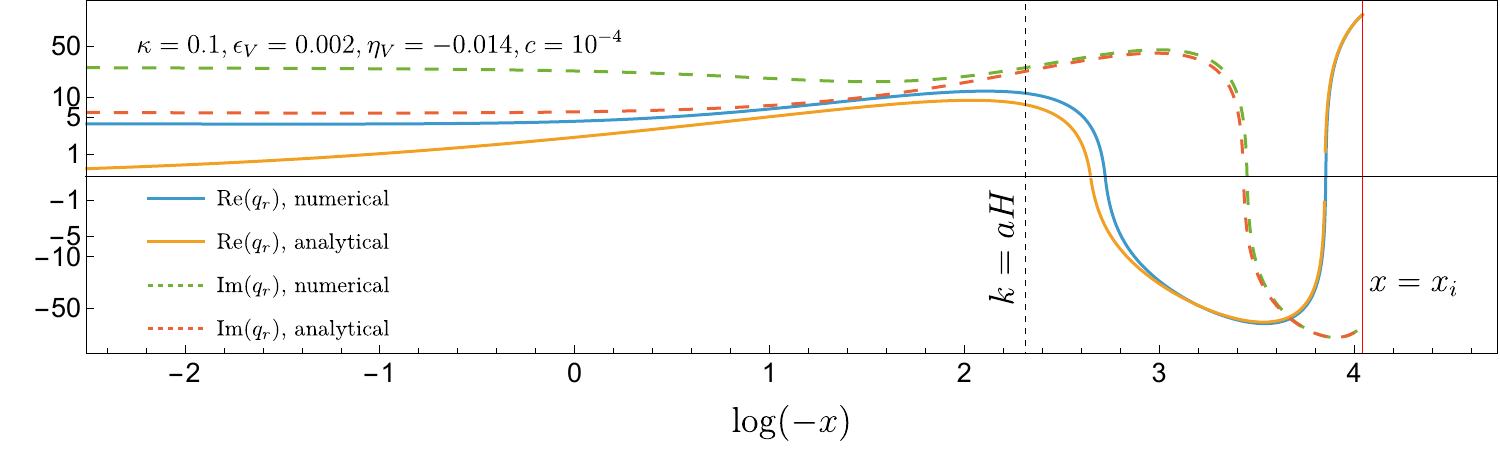}
\:\:
\includegraphics[scale=0.53]{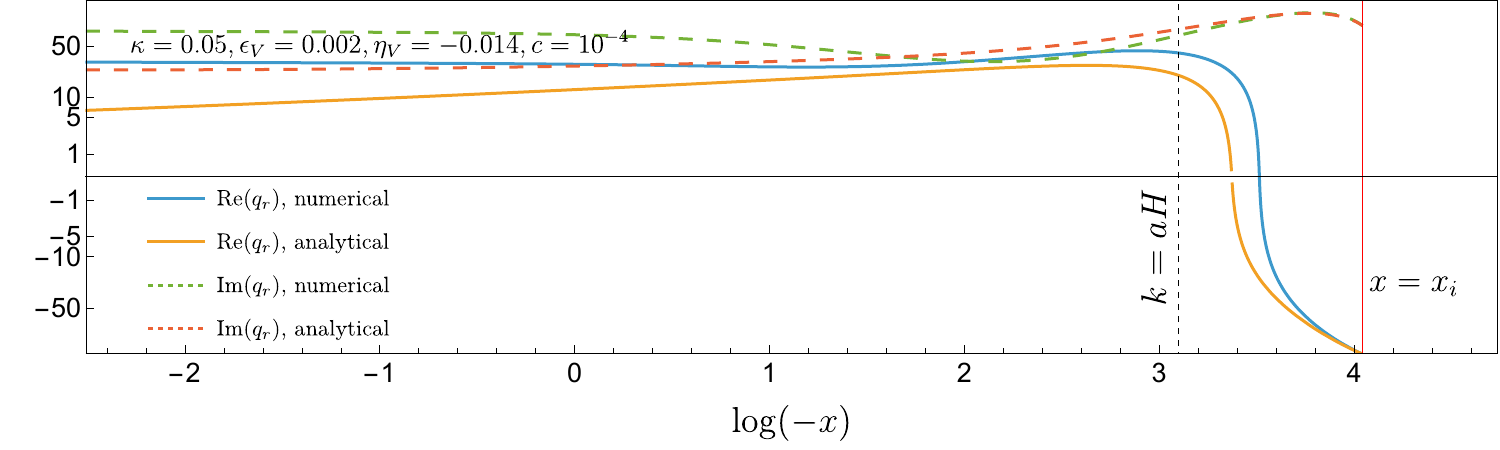}
\caption{\small{A comparison between the analytical and numerical solution for $q_r$ (real parts are shown by solid lines and imaginary parts by dotted lines). The initial conditions are set at $x = x_i = -1/\sqrt{3c} \approx -57$, indicated by the vertical red line. Top to bottom show the results for different values of $\kappa$, in decreasing order. Vertical dashed black lines show when the mode crosses the horizon. Numerical and analytical results agree well for large $\kappa$ but start to deviate for small $\kappa$. Numerical values of $\epsilon_V, \eta_V, c$ are indicated on the plots.}}
\label{fig:qr-qs-numerical-vs-analytical}
\end{figure}

The equations for $q_r, q_s$ are coupled with $x$ dependent coefficients and in general do not admit an analytical solution. If these coefficients are changing slowly, however,
we can replace them by their value at Hubble crossing (for a given $k$), diagonalize the resulting equations and solve them analytically. This is the procedure carried out in ref.~\cite{Lalak:2007vi}. We find that this is a good procedure for large $k$, as shown by an agreement of the analytical and numerical solutions (see fig.~\ref{fig:qr-qs-numerical-vs-analytical}), but there are deviations between the numerical and analytical results at small $k$.

We define $\eta_{ab} = V_{ab}(\Pi, \sigma)/3H^2$, where $V_{ab}$ is the second derivative of the potential along the field direction $a,b$. Taking $a, b$ to be the adiabatic and the orthogonal entropy directions $r, s$ we get
\begin{align}
    \eta_{rr} &= \frac{1}{3H^2}\left(\cos^2 \theta \, \partial_\Pi^2 V + \sech^2\left(\frac{\Pi}{\sqrt{6}M_4}\right) \sin^2 \theta\, \partial_\sigma^2 V + \sech\left(\frac{\Pi}{\sqrt{6}M_4}\right) \sin 2\theta\, \partial_\sigma\partial_\Pi V\right)
    \:,  \\
    \eta_{ss} &= \frac{1}{3H^2}\left(\sin^2 \theta \, \partial_\Pi^2 V + \sech^2\left(\frac{\Pi}{\sqrt{6}M_4}\right) \cos^2 \theta\, \partial_\sigma^2 V - \sech\left(\frac{\Pi}{\sqrt{6}M_4}\right) \sin 2\theta\, \partial_\sigma\partial_\Pi V\right)
    \:,  \\
    \eta_{rs} &= 
    \frac{1}{3H^2}\left(-\frac12\sin 2\theta  \, \partial_\Pi^2 V + \sech^2\left(\frac{\Pi}{\sqrt{6}M_4}\right) \frac12 \sin 2\theta\, \partial_\sigma^2 V + \sech\left(\frac{\Pi}{\sqrt{6}M_4}\right) \cos 2\theta\, \partial_\sigma\partial_\Pi V\right)\:.
\end{align}
The parameters that go into determining the mode function solutions are
\begin{align}
    \xi &= \frac{1}{\sqrt{6}} \tanh\left(\frac{\Pi}{\sqrt{6}M_4}\right)\frac{\dot{X}}{M_4 H}
     \:,\\
    \Theta &= \frac12 \tan^{-1}\left(\frac{2(\eta_{rs}-\xi \sin^3\theta)}{\eta_{rr} -\eta_{ss}-\frac{\dot{X}^2}{M_4^2 H^2}+\xi\cos\theta(1+2\sin^2\theta)}\right)\:,\\
    \label{eq:Theta}
    \mu_{1} &= \sqrt{\frac94 + \lambda_1^2}\:,\:\: \mu_{2} = \sqrt{\frac94 + \lambda_2^2}\:,\\
    \lambda_1^2 &= \frac{3\dot{X}^2}{M_4^2 H^2} - \frac32 \left(\eta_{rr}+\eta_{ss}-\xi\cos\theta\right) - 3 \csc 2\Theta\left(\eta_{rs}-\xi \sin^3\theta\right)\:,
     \\
    \lambda_2^2 &= \frac{3\dot{X}^2}{M_4^2 H^2} - \frac32 \left(\eta_{rr}+\eta_{ss}-\xi\cos\theta\right) + 3 \csc 2\Theta\left(\eta_{rs}-\xi \sin^3\theta\right)\:.
\end{align}
The above quantities are all time-dependent. In places where we evaluate them at horizon-crossing, $k=aH$, for a given $k$-mode, we denote them with a `$*$' in the subscript.

It is useful to know the late time ($\eta\to0$) limit of these parameters, to be able to compare later to the single field case. At late times, $\Pi$ slows down and approaches zero (from eq.~\eqref{eq:pi-bkg-sol}), so we have
\begin{align}
    &\xi\to 0 \: ,\:\:
    \partial_\Pi V \propto \sinh (\Pi/(\sqrt{6}M_4))\to0 \: ,\:\:
    \cos\theta\to 0\, .
    \label{eq:late_limits}
\end{align}
This implies that $\eta_{rs}$ and $\Theta$ also vanish at late times. Further, an explicit computation shows that in this limit, $\mu_1\to3/2+3\epsilon_V-\eta_V$ and $\mu_2\to1/2+7\epsilon_V$. As we show in more detail below, this means that the late time solution for the curvature perturbation will match the single-field result.

The solution for $q_r, q_s$, satisfying the condition~\eqref{eq:qr-qs-ic}, is given as
\begin{align}
    q_r(x) &= \frac{e^{-i\kappa x_i}}{a(x)\,\sqrt{2\kappa}}\sqrt{\frac{x}{x_i}}
    \left(
    \frac{
    \cos \Theta_* e^{i\mu_{1*}\pi/2} H_{\mu_{1*}}^{(1)}(-\kappa x) 
    - \sin \Theta_* e^{i\mu_{2*}\pi/2} H_{\mu_{2*}}^{(1)}(-\kappa x) 
    }{\cos \Theta_* e^{i\mu_{1*}\pi/2} H_{\mu_{1*}}^{(1)}(-\kappa x_i) 
    - \sin \Theta_* e^{i\mu_{2*}\pi/2} H_{\mu_{2*}}^{(1)}(-\kappa x_i)}
    \right)\:,
    \nonumber
    \\
    q_s(x) &=
    \frac{e^{-i\kappa x_i}}{a(x)\,\sqrt{2\kappa}}\sqrt{\frac{x}{x_i}}
    \left(
    \frac{
    \sin \Theta_* e^{i\mu_{1*}\pi/2} H_{\mu_{1*}}^{(1)}(-\kappa x{\phantom{i}}) 
    + \cos \Theta_* e^{i\mu_{2*}\pi/2} H_{\mu_{2*}}^{(1)}(-\kappa x{\phantom{i}}) 
    }{\sin \Theta_* e^{i\mu_{1*}\pi/2} H_{\mu_{1*}}^{(1)}(-\kappa x_i) 
    + \cos \Theta_* e^{i\mu_{2*}\pi/2} H_{\mu_{2*}}^{(1)}(-\kappa x_i)}
    \right)\:.
    \label{eq:qr-qs-analytical-sol}
\end{align}
It is straightforward to see that at $x=x_i$, the expressions for $q_r, q_s$ simply to that in eq.~\eqref{eq:qr-qs-ic}, as expected. In the expressions for $q_r, q_s$ in eq.~\eqref{eq:qr-qs-analytical-sol}, the quantities $\Theta_*, \mu_{1*}, \mu_{2*}$ are evaluated at the Hubble crossing $k = a H$, which translates to 
\begin{align}
    \kappa(x) = \frac{1}{-x\,(1-c\,x^2)}\left(1+\epsilon_V\frac{\left(1-c\,x^2\right)^3}{\left(1-\frac23 c\,x^2\right)^2\left(1+4c\,x^2\right)}\right)\:. 
    \label{eq:kappa-vs-x-HorizonExit}
\end{align}
While the relation between $\kappa$ and $x$ is in general multi-valued, restricting to $x > -1/\sqrt{3c}$ gives a unique relation between $\kappa$ and $x$, as seen from figure~\ref{fig:comovingHubbleRadius}.

At this point we can compare the analytical expression for $q_r, q_s$ from eq.~\eqref{eq:qr-qs-analytical-sol} with the numerical solution obtained by solving the coupled differential equations in.~\eqref{eq:qr-qs-eqn} with the initial conditions in eq.~\eqref{eq:qr-qs-ic}. Figure~\ref{fig:qr-qs-numerical-vs-analytical} shows the comparison for some values of $\epsilon_V, \eta_V, c$, and for a few choices of $\kappa$. We have taken the initial value $x_i = -1/\sqrt{3c}$, the earliest we can allow for an inflating background. This is shown by the vertical red line. We have chosen three representative values for $\kappa$ that correspond to modes that cross the horizon further and further away from the initial condition. The time at which the given mode crosses the horizon is shown by a black vertical dotted line. We see good agreement for modes that cross the horizon away from the initial condition, but the agreement is poor for modes that cross the horizon close to the initial condition. The modes for which the numerical and analytical results don't agree well are the ones that exit the horizon earliest during inflation, and correspond to long wavelength modes entering the horizon at late times. 

\section{Adiabatic Power Spectrum and the CMB}
\label{sec:adiabatic-and-cmb}

In this section we calculate the adiabatic power spectrum, using the numerical and analytical results derived in the previous section. 
We find that the radion fluctuations are the dominant contributor to the adiabatic power spectrum at early times, which leads to significant deviations from an approximately scale-invariant spectrum on large scales. At late times during inflation, the inflaton fluctuations are responsible for generating the adiabatic curvature perturbation and the power spectrum matches the result for single-field inflation. Using numerical fits to the power spectrum, we quantify how the power spectrum deviates from the single-field expectation.

\subsection{Adiabatic power spectrum}
The gauge invariant comoving curvature perturbation $\mathcal{R}$ is given in terms of $Q_r$ as
\begin{align}
    \mathcal{R} = \frac{H}{\dot{X}}Q_r\:.
\end{align}
The dimensionless power spectrum $\Delta_\mathcal{R}^2$ is given as 
\begin{align}
    \Delta_\mathcal{R}^2 &= \frac{k^3}{2\pi^2}\frac{H^2}{\dot{X}^2}\left|Q_r\right|^2
    =\frac{k^3}{2\pi^2}\frac{H^2}{\dot{X}^2} \frac{\left|q_r\right|^2}{\Hsigma}\:.
\end{align}
Using the analytical expression for $q_r$ from eq.~\eqref{eq:qr-qs-analytical-sol}, and using eqns.~\eqref{eq:X},~\eqref{eq:Hubble-slow-roll-corrected}, we can get an analytical expression for the power spectrum. Keeping terms up to linear order in the small parameters, the expression simplifies to
\begin{align}
    \Delta_\mathcal{R}^2(k^2, x)
    =\frac{1}{8\pi^2}\left(\frac{k}{M_4}\right)^2\frac{1}{\epsilon_V + 3 c x_*^2}\left(\frac{x^3}{x_i}\right)
    \left|\frac{
    \cos \Theta_* e^{i\mu_{1*}\pi/2} H_{\mu_{1*}}^{(1)}(-\kappa x) 
    - \sin \Theta_* e^{i\mu_{2*}\pi/2} H_{\mu_{2*}}^{(1)}(-\kappa x) 
    }{\cos \Theta_* e^{i\mu_{1*}\pi/2} H_{\mu_{1*}}^{(1)}(-\kappa x_i) 
    - \sin \Theta_* e^{i\mu_{2*}\pi/2} H_{\mu_{2*}}^{(1)}(-\kappa x_i)}\right|^2\:,
\end{align}
where $x_i$ is the start of inflation, $x_*$ is when the given $k$ exits horizon (see eq.~\eqref{eq:kappa-vs-x-HorizonExit}) and $x$ is unspecified as yet.

To connect to present day observables, we need to evaluate $\Delta_\mathcal{R}^2(k^2, x)$ at the end of inflation. For single-field inflationary models, $\mathcal{R}$ is effectively frozen for super horizon-modes, i.e. $\Delta_\mathcal{R}^2(k^2, x)$ does not change once $x\lesssim x_*$ and one can evaluate it close to $x_*$. For the multi-field case, this is not necessarily true because the isocurvature perturbation can source the curvature perturbation even outside the horizon. However, if the effective mass of the isocurvature fluctuation is not suppressed, the evolution of $\mathcal{R}$ at superhorizon scales is still approximately frozen. This is because the isocurvature fluctuations are massive, and decay quickly once they become superhorizon. 

Therefore, for the present case the evolution of $\mathcal{R}$ is still frozen for modes that are super-horizon: for a given $k$, $\Delta_\mathcal{R}(k^2, x)$ does not change once $x\lesssim x_*$.  
However, to be safe, we will evaluate the expression for $\Delta_\mathcal{R}^2 (k^2, x)$ at $x \ll x_*$. We suppress the second argument of $\Delta_\mathcal{R}^2(k^2, x)$ when not relevant.

The $k$ dependence of $\Delta^2_\mathcal{R}$ can be understood analytically in the early and late times, and provides a useful check of the numerical results. At early times, we know that the adiabatic fluctuations are closely aligned with the radion fluctuations. Purely radion fluctuations are expected to scale as $k^{3-\sqrt{9-4m^2/H^2}}$ where $m^2/H^2$ is the mass squared of the radion in units of Hubble. From eq.~\eqref{eq:mass-cPiPi} we know that the radion mass changes as a function of $c \Hsigma^2 \eta^2$. Close to the start of inflation, when $c \Hsigma^2\eta^2\lesssim1/3$, the radion fluctuations have a mass squared of $m^2 \sim (4/3) H^2$, which translates approximately to a $k^{1}$ scaling. At later times, when $c \Hsigma^2 \eta^2 \ll 1$, $m^2/H^2 \to 2$ and we expect a $k^2$ scaling for the radion fluctuations. We therefore expect a \textit{blue-tilted} spectrum at early times, with a varying tilt as the component of the adiabatic direction along the radion fluctuations changes, before the inflaton takes over and the spectrum becomes red-tilted. 

At late times, close to the end of inflation, the dynamics reduces to that of a single-field model. The modes that exit during this period correspond to large $k$, and in this limit, $\Theta_* \to 0,\: \mu_{1*} \to 3/2 + 3\epsilon_V -\eta_V$ (see discussion around eq.~\eqref{eq:late_limits}). For these $k$, the power spectrum simplifies to
\begin{align}
    \Delta_\mathcal{R}^2(k^2, x) \:\: \underset{\text{large $k$}}{=}\:\:& 
    \frac{1}{8\pi^2}\left(\frac{k}{M_4}\right)^2\frac{1}{\epsilon_V + 3 c x_*^2}\left(\frac{x^3}{x_i}\right)
    \left|\frac{H_{\mu_{1*}}^{(1)}(-\kappa x)}{H_{\mu_{1*}}^{(1)}(-\kappa x_i)}\right|^2\:.
\end{align}
For $-\kappa x \ll 1$ and $-\kappa x_i \gg 1$, we can use the asymptotic form of Hankel function
\begin{align}
    H_{\nu}^{(1)}(-\kappa x)\underset{-\kappa x \to 0}{ = }\:\frac{\Gamma(\nu)}{i\pi}\left(\frac{-\kappa x}{2}\right)^{-\nu}\:,\:\:
    H_{\nu}^{(1)}(-\kappa x)\underset{-\kappa x \to \infty}{ = }\:\sqrt{\frac{2}{\pi(-\kappa x)}}e^{-i\kappa x}\,e^{-i\pi/2(\nu+1/2)}\:,
    \label{eq:Hankel-limits}
\end{align}
to get
\begin{align}
    \Delta_\mathcal{R}^2(k^2) \:\: \underset{\text{large $k$}}{=}\:\:& 
    \frac{1}{8\pi^2}\left(\frac{\Hsigma}{M_4}\right)^2\frac{1}{\epsilon_V + 3 c x_*^2}\left(\frac{k}{k_\text{ref}}\right)^{2\eta_V - 6\epsilon_V}\:,
    \label{eq:Delta-S-large-k}
\end{align}
where we have defined a reference momentum $k_\text{ref}$.
We see that the power spectrum matches the result for single field case for $3 c x_*^2 \ll \epsilon_V$. The spectral tilt in this limit reduces to the expected result of $2\eta_V - 6\epsilon_V$.

\begin{figure}[h]
\centering
\includegraphics[scale=0.65]{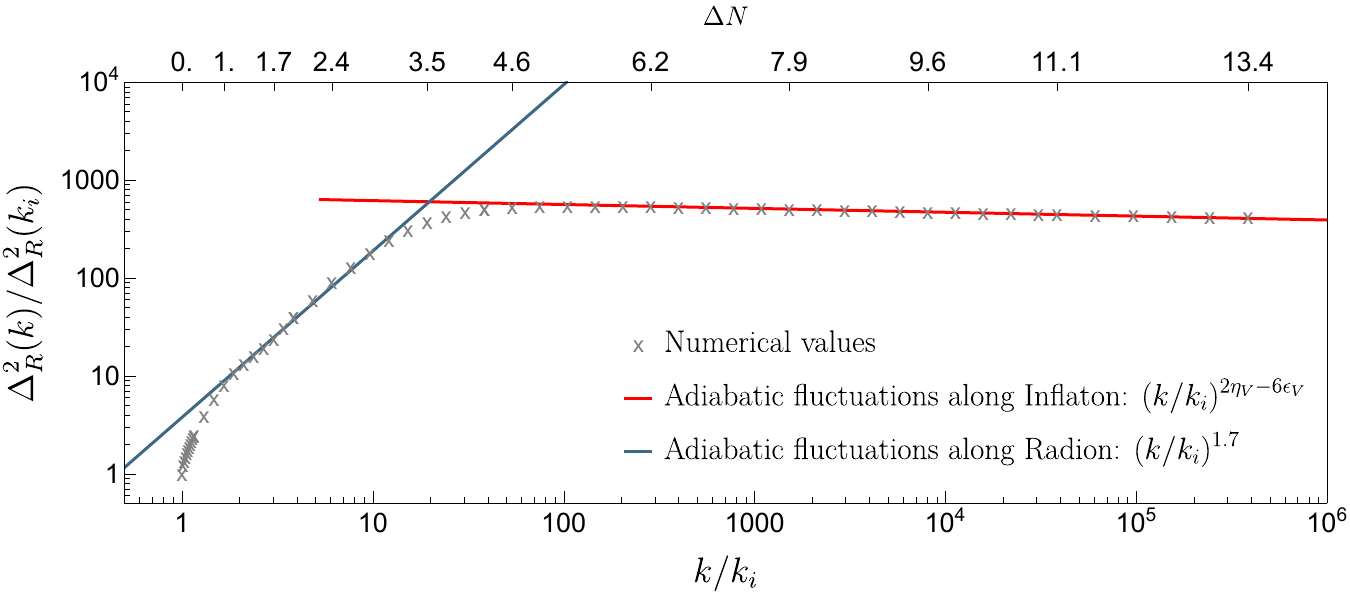}
\caption{
\small{Rescaled power spectrum of the adiabatic scalar fluctuations as a function of $k/k_i$. The parameters chosen are $\epsilon_V = 0.002, \eta_V = -0.014, c = 10^{-4}, x_f=-10^{-25}$ and $x_i = -1/\sqrt{3c} \approx -57$. Numerical results are shown by gray crosses. The early time expectation from being aligned along the radion direction is shown in dark blue (a scaling of $k^{1.7}$ is seen to fit well), while the late time expectation from the single-field limit (with the same $\epsilon_V, \eta_V$ as here) is shown in red ($k^{2\eta_V - 6\epsilon_V}$ scaling). The upper x-axis shows the number of e-folds counted from the start of inflation. For smaller $k$, a significant deviation is observed and the power spectrum is reduced.
}}
\label{fig:power-spectrum}
\end{figure}

We now present the numerical result for the power spectrum as a function of $k/k_i$ where $k_i = a(x_i)H(x_i)$ is the smallest $k$ that we consider (which depends on where we set the initial conditions).
Figure~\ref{fig:power-spectrum} shows the power spectrum, normalized to the power spectrum evaluated at $k_i$, $\Delta_\mathcal{R}^2(k_i)$, for the representative choice of parameters ($\epsilon_V = 0.002, \eta_V = -0.014, c = 10^{-4}, x_f=-10^{-25}$) and for $x_i = -1/\sqrt{3c}$.
Also shown are the early time best-fit $k$ scalings from alignment along the radion direction, and the late time expectation from the single-field limit. The top $x$-axis shows the number of e-folds counted from the start of inflation, and is a measure of how much time a given mode has spent within the horizon.
We see that for small and large $k$, the slope of the power spectrum is consistent with the expected limits. Crucially, for smaller $k$, the power spectrum is suppressed relative to the single-field case. 

\begin{figure}[h]
\centering
\includegraphics[scale=0.63]{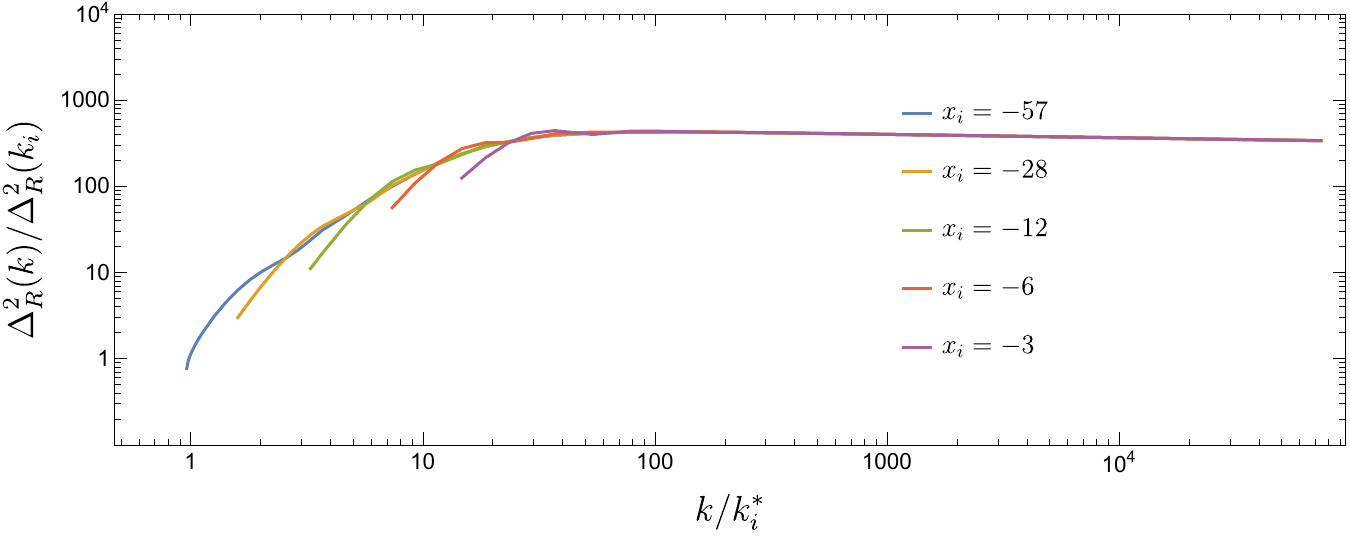}
\caption{\small{Rescaled power spectrum of the adiabatic scalar fluctuations as a function of $k/k_i^*$ for a few choices of the initial time $x_i$, where $k_i^*$ is the smallest $k$ for the smallest $x_i$. The fixed parameters are $\epsilon_V = 0.002, \eta_V = -0.014, c = 10^{-4}$. Early blue tilt followed by a late red tilt persists as in fig.~\ref{fig:power-spectrum}. The power spectra are rescaled to match at late times.}}
\label{fig:power-spectrum-vary-xi}
\end{figure}

So far we have imposed the boundary condition at $x_i = -1/\sqrt{3c}$. We now relax that assumption and consider different choices of $x_i$ (all larger than $-1/\sqrt{3c}$). Figure~\ref{fig:power-spectrum-vary-xi} shows the numerically computed power spectrum as a function of $k/k_i^*$ for $\epsilon_V = 0.002, \eta_V = -0.014, c = 10^{-4}$ but for different choices of $x_i$ (here $k_i^*  = a(x_i)H(x_i)$ evaluated at $x_i =-1/\sqrt{3c}$). We see that the general feature of an initial blue tilt followed by a late red tilt persists. Note that for two different choices of the initial time, $x_i^{(1)}$ and $x_i^{(2)}$, satisfying $x_i^{(1)}  > x_i^{(2)}$, the corresponding smallest mode that experiences inflation, denoted by $k_i^{(1)}, k_i^{(2)}$, satisfy $k_i^{(1)} > k_i^{(2)}$. As a result, lines with larger $x_i$ start at larger values of $k/k_i^*$ in fig.~\ref{fig:power-spectrum-vary-xi}. 
Further, for a fixed value of $x_f$ (the end of inflation), as $x_i$ increases the amount of time during inflation where the dynamics deviates from single-field inflation decreases. As a result, the range of $k/k_i^*$ for which the spectrum is blue-tilted also reduces. Therefore, for larger $x_i$, the Universe spends less time in the radion-dominated period where we observe significant deviations from single-field inflation.

In summary, due to the presence of additional radion dynamics, there are two differences in the power spectrum compared to the single-field case:
\begin{itemize}
    \item Amplitude: the power spectrum is suppressed at smaller $k$ values.
    \item Tilt: the tilt is not constant, and is significantly modified at low $k$, making the power spectrum go from red-tilted at large $k$ values to blue-tilted at small $k$ values. These limits can be understood by the dynamics being dominated by the radion at early times, and by the inflaton at late times.
\end{itemize}
At larger scales we see that the power spectrum is smaller than the slow-roll expectation. Such a feature is in fact preferred by the current CMB data, e.g. as discussed in ref.~\cite{Petretti:2024mjy}. This dip in power spectrum is a direct consequence of the radion fluctuations dominating the adiabatic modes at early times during inflation. To establish how well this fits the data, we need to perform a fit of the shape of the power spectrum to data. In what follows, we quantify this using fit functions to the numerical power spectrum. A detailed treatment will be left for future work.

\subsection{CMB}
\label{sec:CMB}
In this subsection we discuss the implications of the observed CMB data for our scenario, given the prediction of the shape of the adiabatic power spectrum. Using single-field inflation as the benchmark, we quantify the deviation in the power spectrum from the single-field expectation, as a function of the angular multipole $\ell$. For this, it is useful to have an analytical expression for the power spectrum as a function of $k$ and other parameters. Since the analysis was numerical, we use a fit to the numerical data.

We know from the results earlier that the large $k$ amplitude and tilt of the adiabatic power spectrum are consistent with the single-field expectation of a constant red-tilt, and there is a blue tilt for smaller $k$ values. With this in mind, a useful parameterization of the power spectrum is 
\begin{align}
    \log (\Delta_\mathcal{R}^2(k)/\Delta_\mathcal{R}^2(k_i)) = a_1 + a_2 \log \left(k/k_i\right) + a_3 \tanh \left(\frac{\log\left(k/k_i\right)-a_4}{a_{5}}\right)\:,
    \label{eq:fit-function}
\end{align}
where $a_i, i = 1, \cdots 5$ are constants and we have chosen to normalize both $k$ and $\Delta_\mathcal{R}^2(k)$ by their values at the earliest time during inflation, $k_i$ and $\Delta_\mathcal{R}^2(k_i)$ respectively. The earliest time corresponds to $x = x_i = -\eta_i \Hsigma$, where $\eta$ is the conformal time and $\Hsigma$ is the late time Hubble constant during inflation. The functional form is such that it approaches the single-field result of a linear function of $\log \left(k/k_i\right)$ when $a_3 \to 0$ or when $\log\left(k/k_i\right) \ll a_4$. Therefore in our parameterization, $a_4$ indicates at what value of $\log (k/k_i)$ the shape starts to differ from the single field expectation, $a_3$ indicates the size of this deviation and $a_5$ indicates how fast is the transition. The constants $a_i$ are determined by a fit to the numerical data, and will be a function of all the other parameters that go into the calculation.  

\begin{figure}[h]
\centering
\includegraphics[scale=0.7]{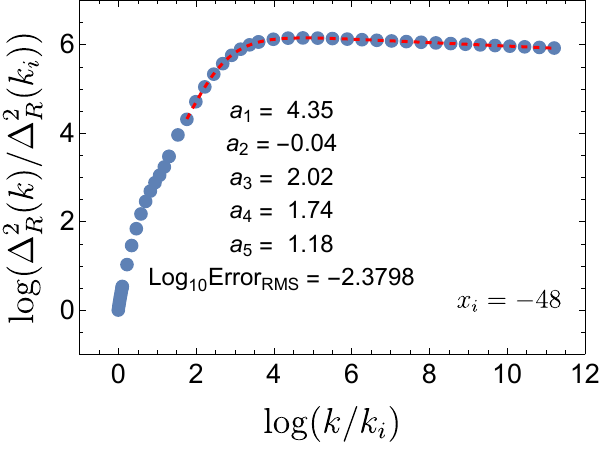}
\qquad
\includegraphics[scale=0.7]{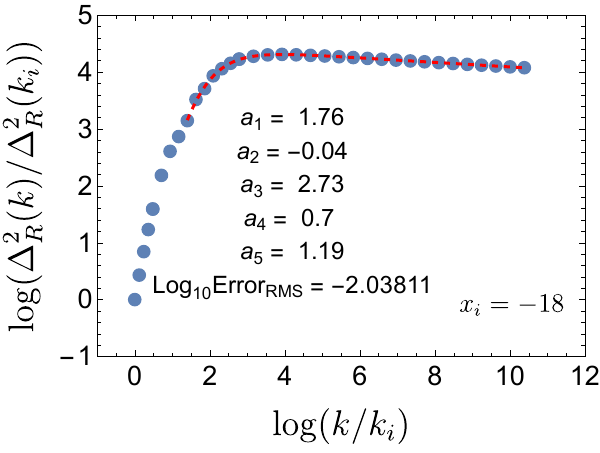}
\caption{\small{Adiabatic power spectrum for two different values of $x_i = -\eta_i \Hsigma$ ($\Hsigma$ is the late time Hubble constant during inflation), for fixed $\epsilon_V = 0.002, \eta_V = -0.014$. Numerical values are shown by blue dots, and the fit is shown by the red dashed curve. The end points of the fit curve indicate the range of values for which the fit is performed. The fit has the functional form as in eq.~\eqref{eq:fit-function}. The best fit values for the parameters $a_1, a_2, a_3, a_4, a_5$ are shown on the plots, along with the root mean square error of the fit.}}
\label{fig:power-spectrum-adiabatic-fits}
\end{figure}

Figure~\ref{fig:power-spectrum-adiabatic-fits} shows the numerical results and the fit function for two representative values of $x_i$, for $\eta_V = -0.014, \epsilon_V=0.002$ (chosen to be consistent with the tilt of $0.96$ for large $k$). The numerical results are in blue dots, and the fit function is in red dashed. To have a high fidelity of the fit, we have restricted to fit only in a range of values of $k/k_i$ which includes the full red tilted part and a significant portion of the initial blue tilted part (including the full blue tilted part is ruled out anyway due to CMB constraints). The range in which the fit is performed is indicated by the end points of the red dashed line. The fit quality is high, as indicated by the root mean square error of the fit, which is always less than $10^{-2}$ for the parameter choices considered (see table~\ref{tab:fit-coeffs}). We see that for different values of $x_i$, the shape is qualitatively similar, although there are slight changes in the values of the best-fit parameters. As a consistency check, the best-fit value for the constant $a_2$ is always $2\eta_V - 6\epsilon_V = -0.04$, as it should be to be consistent with the single-field result for large $k$. The value of the best fit parameters for a few values of $x_i$, for fixed $\epsilon_V, \eta_V$, along with the root mean square error of the fit to the numerical data is presented in table~\ref{tab:fit-coeffs}. It is straightforward to obtain the best-fit values for other choices of the underlying parameters, and can be provided on request.

\begin{table}[h!]
\centering
\begin{tabular}{|| c | c | c | c | c | c | c ||}
\hline
$x_i$ & $a_1$ & $a_2$ & $a_3$ & $a_4$ & $a_5$ & $\log_{10}$(RMS Fit Error) \\
\hline\hline
-57 & 4.08 & -0.04 & 2.18 & 1.67 & 1.23 & -2.15\\
 \hline
-48 & 4.39 & -0.04 & 1.95 & 1.77 & 1.17 & -2.36\\
 \hline
-38 & 3.67 & -0.04 & 1.97 & 1.55 & 1.16 & -2.55\\
 \hline
-28 & 2.96 & -0.04 & 2.19 & 1.19 & 1.21 & -2.02\\
 \hline
-18 & 1.76 & -0.04 & 2.73 & 0.7 & 1.19 & -2.04\\
 \hline
\hline
\end{tabular}
\caption{\small{Best fit values of parameters in the functional form for the adiabatic power spectrum (see eq.~\eqref{eq:fit-function}) for a few values of $x_i$, for fixed $\epsilon_V = 0.002, \eta_V = -0.014$.}}
\label{tab:fit-coeffs}
\end{table}

Given the fit coefficients, one needs further information to uniquely fix where the CMB data constrains the shape. The CMB window, for $k$ values in $(10^{-4} - 10^{-1}) \text{Mpc}^{-1}$, spans three orders of magnitude, and corresponds to a certain number of e-folds, $N_\text{CMB}$, before the end of inflation. $N_\text{CMB}$ depends on the reheat temperature and further details of reheating, but is approximately around $N_\text{CMB} \sim 40-60$. To be consistent with the standard paradigm of inflation we need to have at least $N_\text{CMB}$ e-folds of inflation in our scenario. While we have talked so far about the start of inflation, at $x = x_i$, to uniquely fix the scenario, we also need to specify the end of inflation, at $x = x_f$. The total number of e-folds $N_\text{total} = \log (a(x_f)/a(x_i))$ should be larger than $N_\text{CMB}$, and this can be easily arranged by choosing $x_f$ appropriately. Given $x_f$, or equivalently $N_\text{total}$, this determines the largest $k$-mode, $k_\text{max}$, that experiences inflation, and therefore the maximum value of the ratio $k/k_i$ for modes which undergo inflation. To identify the CMB window we then need to determine the value $x_{\text{CMB}}$, which satisfies
\begin{align}
    a(x_\text{CMB})/a(x_f) = e^{-N_\text{CMB}}\:,
    \label{eq:xf-vs-CMB-window}
\end{align} 
and then calculate the $k$ value of the mode that exited the horizon around $x_\text{CMB}$ (see eq.~\eqref{eq:kappa-vs-x-HorizonExit}). The CMB window spans the three orders of magnitude before this $k$ value (taking the right end of the window to correspond to $N_\text{CMB}$ e-folds before the end of inflation). 

Effectively, the CMB window can be slid across the x-axis in fig.~\ref{fig:power-spectrum} or fig.~\ref{fig:power-spectrum-vary-xi}, and each location of the window corresponds to a different value of $x_f$. The left-most point of the window corresponds to $k = 10^{-4} \text{ Mpc}^{-1}$ and the right most point corresponds to $k = 10^{-1} \text{ Mpc}^{-1}$. Once the CMB window is fixed, one can rescale the value of $\Delta^2_\mathcal{R}$ to equal the amplitude $A_s = 2.1\times 10^{-9}$, at the pivot scale $0.05 \text{ Mpc}^{-1}$, and then calculate how well the shape of the power spectrum fits the CMB data. Further, once the CMB window is fixed, one can identify the locations corresponding to the angular multipole $\ell$, using $\ell\approx k \times 1.4\times 10^{4} \text{ Mpc}$.

\begin{figure}[h!]
\centering
\includegraphics[scale=0.5]{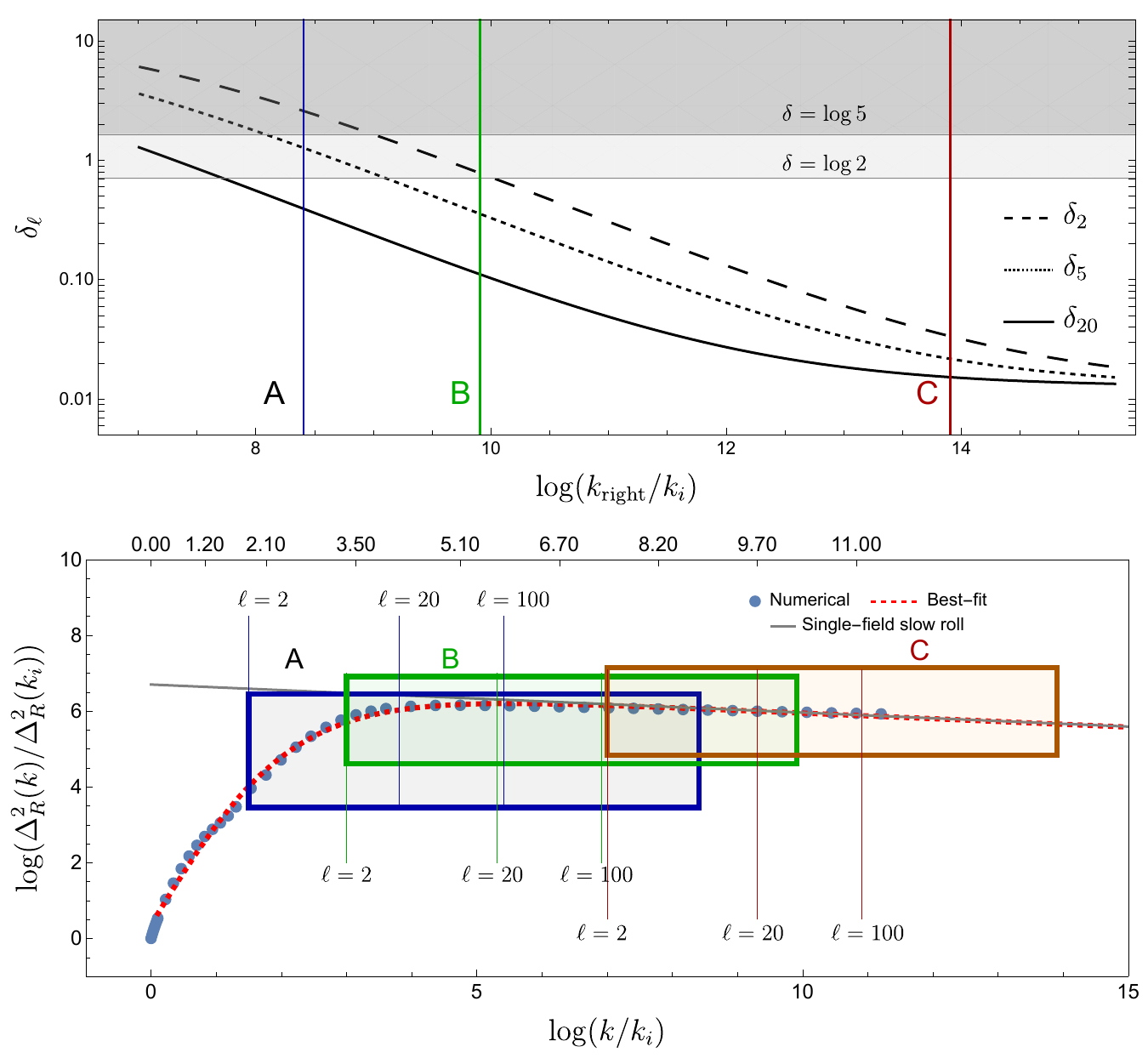}
\caption{\small{Top: Deviation from single-field inflation parameterized by $\delta_\ell$ (see eq.~\eqref{eq:delta-l}), for $\ell=2, 5, 20$, as a function of the right end of the CMB window, $\log(k_\text{right}/k_i)$. Different points in the plot correspond to different locations of the CMB window, and amount to different choice of $x_f$ (as defined by eq.~\eqref{eq:xf-vs-CMB-window}, for a fixed $N_\text{CMB}$). Bottom: Numerical data (blue dots), best fit (red dashed) and single field result (gray solid) for the adiabatic power spectrum. Three different choices of CMB window are shown. The three choices for the CMB window in the bottom panel are indicated by vertical lines in the top panel.}}
\label{fig:power-spectrum-adiabatic-slidingCMBwindow}
\end{figure}

The bottom panel of fig.~\ref{fig:power-spectrum-adiabatic-slidingCMBwindow} shows three choices of the CMB window, superimposed on the adiabatic power spectrum. Both numerical data (blue dots) and the fit (red dashed) are shown, along with the constant red-tilt line from single-field limit (in gray). Vertical lines indicate the location of the angular multipoles for $\ell = 2, 20, 100$. It is clear that in scenario $C$, the power spectrum is red-tilted for the entire range, while in scenarios $A$ and $B$, the power spectrum has blue tilt at lower angular multipoles, to varying degrees. The three choices for the CMB window correspond to different choices for when the inflation ends, parameterized by $x_f = -\eta_f \Hsigma$, keeping all other parameters fixed. Taking the right end of the window to correspond to $N_\text{CMB}$ e-folds before the end of inflation, we can calculate the value of $x_f$ for a given choice of the CMB window, using eq.~\eqref{eq:xf-vs-CMB-window}. For example, for $N_\text{CMB} = 60$, the three CMB windows indicated by $A,B,C$ in the top panel of fig.~\ref{fig:power-spectrum-adiabatic-slidingCMBwindow} correspond to $x_f \approx -6.3\times 10^{-29}, -2\times 10^{-29}, -4.4\times 10^{-30}$ respectively.

To quantify the deviation at small $\ell$ values, we first obtain the single-field result, $\widetilde{\Delta}_\mathcal{R}^2$ (by setting $c=0$), which only fits the red-tilt at large $k/k_i$ values, and then quantify the deviation from the single-field result for a given $\ell$ by $\delta_\ell$, defined as
\begin{align}
    \delta_\ell = 
    \left|
    \log \left(\frac{\Delta^2_\mathcal{R}(k)}{\widetilde{\Delta}^2_{\mathcal{R}}(k)}\right)
    \right|_{k = \ell/(1.4 \times 10^4 \text{ Mpc})}\:.
    \label{eq:delta-l}
\end{align}
As we vary the CMB window, or equivalently change $x_f$, we can calculate $\delta_\ell$ and see how the deviation at different $\ell$ values change. 

The top panel of fig.~\ref{fig:power-spectrum-adiabatic-slidingCMBwindow} shows $\delta_2, \delta_5, \delta_{20}$ as a function of the right end of the CMB window $\log(k_\text{right}/k_i)$, as the CMB window is moved. Values corresponding to scenario $A, B, C$ from the top panel are indicated in the bottom panel. We see a monotonic growth in $\delta_\ell$ as we make the CMB window cover the blue-tilted region more, and the order $\delta_2 > \delta_5 > \delta_{20}$ is preserved. We further see that even for a small deviation at $\ell=20$, with $\delta_\ell \sim 0.04$ (i.e. the power spectrum deviates from the single field case by $4\%$), we can have a substantial deviation at $\ell=5,2$, with $\delta_5 = 0.15$ (power spectrum deviates by $13\%$), $\delta_2 = 0.26$ (power spectrum deviates by $30\%$). For larger values of the deviation at $\ell=20$, $\delta_{20}\sim0.13$, we can have a much larger deviation at $\ell=5, 2$. The important point is that the deviation is gradual and not localized at any given $\ell$. 
The low $\ell$ modes of CMB can have a reduced power, as seen by the dip in the data point for smaller $\ell$ and the larger error bars for these data points~\cite{Planck:2018vyg}. As a rough guideline, the $\ell = 2$ value can be a factor of $5$ smaller and $\ell = 5$ can be a factor of $2$ smaller. In fig.~\ref{fig:power-spectrum-adiabatic-slidingCMBwindow} we have indicated regions with $\delta_\ell < \log(5)$ and $\delta_\ell < \log(2)$, which indicates how far to the left the CMB window can be pushed while being consistent with data.

One can fit a function to obtain a functional form for $\delta_\ell$ as a function of $\log(k_\text{right}/k_i)$ for a given choice of model parameters. We find a cubic fit works well in the region which covers scenarios A, B, \& C, and for $\ell = 2, 5, 20$ we get
\begin{align}
    \delta_2 &= 
    63.51 
    -14.77 \log(k_\text{right}/k_i) 
    + 1.14 \log(k_\text{right}/k_i)^2 
    - 0.03 \log(k_\text{right}/k_i)^3\:,
    \\
    \delta_5 &= 
    40.09 
    -9.58 \log(k_\text{right}/k_i) 
    + 0.76 \log(k_\text{right}/k_i)^2 
    - 0.02 \log(k_\text{right}/k_i)^3\:,
    \\
    \delta_{20} &= 
    14.69 
    -3.56 \log(k_\text{right}/k_i) 
    + 0.29 \log(k_\text{right}/k_i)^2 
    - 0.007 \log(k_\text{right}/k_i)^3\:.
    \label{eq:delta-vs-delta-fits}
\end{align}

There are hints of a lower power (compared to the single-field expectation) for smaller values of $\ell$ in the Planck 2018 CMB measurements, which can be potentially explained by the scenario here. It would be interesting to quantify this further and perform a more detailed analysis, which is left to future work.

\subsection{Sensitivity to the start of inflation}
\label{subsec:initial-conditions}
The power spectrum calculated in this section and presented in figure~\ref{fig:power-spectrum-adiabatic-slidingCMBwindow} represents the calculable power spectrum from quantum fluctuations around the inflationary background. If the deviations from single-field expectations on large scales in our model are to be visible, the longest modes of the CMB must have experienced only a few e-folds of inflation (three e-folds for benchmark B above). This means that our predictions are sensitive to the initial conditions at $\eta = \eta_i$, as the Universe has not inflated long enough to wash out the initial conditions. There are two ways in which this sensitivity comes about: the initial condition for background fields and the choice of mode function, which we discuss in turn.

In order to make predictions we have assumed homogeneous background solutions for $\Pi$ and $\sigma$. If we take the initial time slice at $\eta_i$ to be the starting time of the Universe (i.e. there was no cosmological epoch preceding inflation), then we would expect that the initial field profiles are initially highly inhomogeneous with the energy density in these inhomogeneities decaying $\propto 1/a^4$~\cite{East:2015ggf}. This means that the Fourier modes $\Delta \Pi_k (\eta_i)$, $\Delta \sigma_k (\eta_i)$ decay as $e^{-N_i}$, where $N_i$ is the number of e-folds of inflation after $\eta_i$. If we parametrize the initial inhomogeneities by a single scale $\Lambda$, 
\begin{align}
    \Delta \Pi_k (\eta_i)\sim \Delta \sigma_k (\eta_i) \sim \mathcal{O}(\Lambda) \, ,
\end{align}
then the two-point function $\Delta^2_{\rm init} (k) $ due to the initial inhomogeneities after $N$ e-folds of inflation scales as 
\begin{align}
    \frac{\Delta^2_{\rm init} (k) }{\Delta_\mathcal{R}^2(k)} \sim \frac{\Lambda^2 e^{-2N_i}}{H_\sigma^2} \, ,
    \label{eq:PS_initial}
\end{align}
where $\Delta_\mathcal{R}^2(k)$ is the power spectrum from the quantum fluctuations. We know of no way to reliably calculate $\Lambda$, but if we require that $\Delta^2_{\rm init}/\Delta_\mathcal{R}^2 < 1$ at the left end of the CMB window for benchmark~B, for example, then we find that the allowed scale of inhomogeneities should satisfy
\begin{align}
   \frac{\Lambda}{H_\sigma} \lesssim e^{3.5} \simeq 30 \, .
\end{align}

Moving the CMB window to the right would increase $N_i$ in eq.~\eqref{eq:PS_initial} and erase both these inhomogeneities and the features on large scale that we have calculated. An alternative is that our model matches onto a prior cosmological epoch at $\eta = \eta_i$ that led to homogeneous initial conditions. This could occur, for example, in landscape models in which the Universe experiences eternal inflation before tunneling to a homogeneous state where the branes are close together, prior to leading to the dynamics presented here. Finally, it could just be that fluctuations were not so large initially. 

Additionally, the results  depend on the choice of the initial condition imposed at $\eta = \eta_i$ and the choice of $\eta_i$ itself. In principle $\eta_i$ is a free parameter. But since we need to choose the mode functions at this time, we can also use it as a way of deducing the sensitivity to our choice. Figure.~\ref{fig:power-spectrum-vary-xi} shows the power spectrum for a few choice of $\eta_i$ assuming the initial condition at $\eta_i$ is a plane wave. We see that the power spectrum at small $k$ can change by a factor of $\sim 5-10$ as $\eta_i$ is varied in the range we chose. Alternatively, we can parameterize  the initial conditions at $\eta_i$ with a modified form to test sensitivity to our plane wave choice. This is done in appendix~\ref{app:initial-conditions}. We find that different choices modify the overall size of the power spectrum by $\mathcal{O}(1)$ at low $k$, but leave the qualitative shape of the spectrum unchanged.

We conclude that models such as the one we consider with a finite initial time for inflation can lead to measurable variations at small $k$. Such modifications can be consistent with current constraints and might even lead to a better fit to the data. However, due to large uncertainties we cannot necessarily pin down details of the underlying inflationary scenario. We view our predictions as one potential target that might shed light on the early Universe.


\section{Isocurvature and Non-Gaussianities}
\label{sec:isocurvature-and-NG}

In this section we briefly discuss the isocurvature and non-Gaussianities that can be produced in this model.

\subsection{Isocurvature}

Since there is more than one field active during inflation, isocurvature fluctuations are generated during inflation. We already identified the gauge-invariant combination, $q_s$, corresponding to these fluctuations in sec.~\ref{sec:linear-fluctuations}. We also noted that the  isocurvature fluctuations are generated by the radion fluctuations since at late times, the classical field trajectory is mostly along the inflaton (as indicated by the behavior of $\cos\theta$, see fig.~\ref{fig:theta-angle}). 
However, these isocurvature fluctuations do not survive to have visible effects either on the CMB or at smaller scales.  

\begin{figure}[h]
\centering
\includegraphics[scale=0.6]{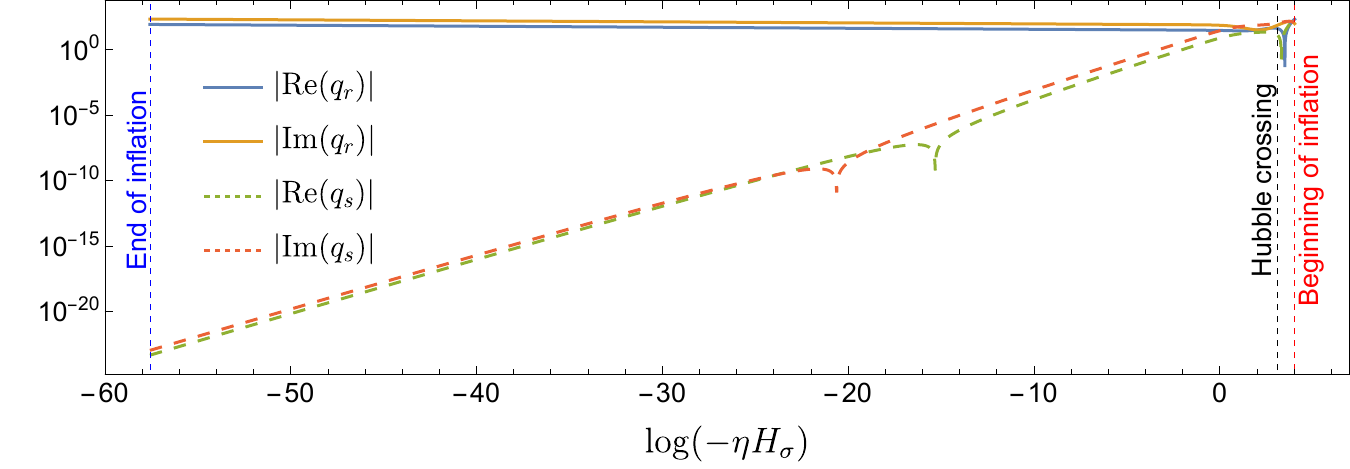}
\caption{\small{Absolute value of the real and imaginary parts of $q_r$ and $q_s$ as a function of $\eta$. Vertical dashed lines indicate the start of inflation (red), Hubble crossing (black) and end of inflation (blue). We see that $|q_s| \ll |q_r|$ very quickly, once the mode crosses the horizon. Here, $\kappa=0.1, \epsilon_V = 0.002, \eta_V=-0.014, c = 10^{-4}, x_i = -1/\sqrt{3c}$.}}
\label{fig:qr-vs-qs}
\end{figure}

The first reason is that the mass of $q_s$ approaches $H$ very quickly and therefore the modes decay once they cross the horizon. This can be seen explicitly by looking at the numerical solution for $q_s$ and comparing it with $q_r$ for example, which does not decay significantly outside the horizon. This is shown clearly in fig.~\ref{fig:qr-vs-qs} for some choice of parameters. Even though the modes have to be further evolved after inflation ends until horizon reentry we do not expect the post-reheating evolution to compensate for this suppression. As a result, the power in $q_s$ when they re-enter the horizon post reheating is very suppressed.   

Furthermore, since the radion couples to other states in the theory, it is expected to decay to them at some point during the cosmological history, as dictated by its decay width. Taking its mass to be order $H$ and couplings to the SM states to be suppressed by $1/M_4$, an estimate for its lifetime is 
\begin{align}
    \Gamma \sim \frac{g_* m_\Pi^3}{8\pi M_4^2} \:,
\end{align}
where $g_*$ is the number of degrees of freedom of the radiation bath.
If we take the radion to be stabilized immediately after inflation ends, then $m_\Pi \sim \Hsigma$, and the radion fluctuations will decay to SM radiation at a temperature $T_{\rm decay}$ when $\Gamma = H$. This temperature is given by
\begin{align}
    T_{\rm decay}\sim \frac{m_\Pi^{3/2}}{ (8\pi M_4)^{1/2}} = \frac{\Hsigma^{3/2}}{ (8\pi M_4)^{1/2}} \:.
\end{align}
For high-scale inflation, $H_\sigma \sim 10^{13}$ GeV, the radion decays at temperatures of $T_{\rm decay} \sim 6\times 10^9$ GeV.
The decay products are expected to thermalize with the Standard Model plasma in the simplest scenario. As a result, no effect of the isocurvature survives to be visible. 

\subsection{Non-Gaussianities}

Being an interacting theory, we generically expect non-gaussianities to be present in our theory. Apart from an inflaton and a 4D graviton, the additional states in the theory are the radion (spin 0) and the KK gravitons (spin 2), which contribute to the three-point function (bispectrum) of $\mathcal{R}$.\footnote{We have integrated out the KK modes in the 4D EFT we have considered. To calculate the non-Gaussianities we would need to include the KK modes in the EFT.} 
In the standard single-field case, the squeezed-limit of the bispectrum can have information about the dynamics. As we have seen, in our model the deviations from single-field results are seen in small $k$ modes. This means that the standard estimate for the bispectrum in the squeezed limit will receive modifications if the smallest $k$ in the bispectrum is sufficiently small. 
Such a modification can be an interesting probe of the dynamics near the start of inflation if it is sufficiently large to be observable. 

\begin{figure}[h]
\centering
\includegraphics[scale=0.5]{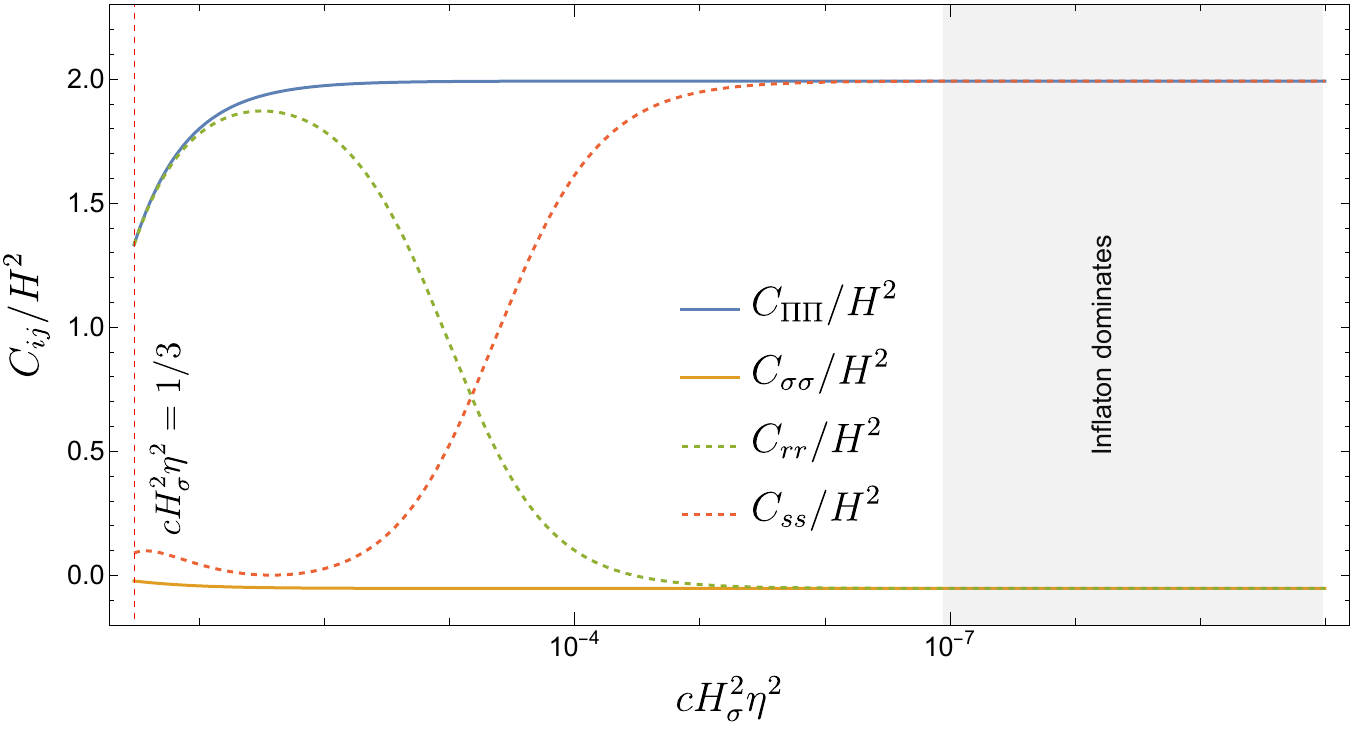}
\caption{\small{Coefficients $C_{ij}$ (appearing in eq.~\eqref{eq:EOM-FluctuationsAlongFields}) in units of Hubble for the inflaton $\sigma$ (orange solid), radion $\Pi$ (blue solid). Also shown are the corresponding quantities for the adiabatic mode (green dashed) and the entropy mode (red dashed), see app.~\ref{app:EOM-fluctuations-along-adiabatic-entropy-direction}. We have taken $\epsilon_V = 0.002, \eta_V = -0.014$.}}
\label{fig:masses-Crr-Css}
\end{figure}

However, we expect the detection of such a modification, even if large,  to be challenging due to the smaller statistics of squeezed configurations for which this effect could be measured.  Another potential issue is that the standard calculation assuming constant mass terms, which we briefly present below, seems to give a small result. 
We  estimate the squeezed limit of the bispectrum leaving the issue of measurability and the time dependence of the mass for further work. At late times, the radion fluctuations have a mass around Hubble (see eq.~\eqref{eq:mass-cPiPi}), and they are approximately orthogonal to the adiabatic fluctuations. The amplitude of the bispectrum of adiabatic fluctuations can be estimated in the squeezed limit $k_\text{long}/k_\text{short} \ll 1$ to be~\cite{Chen:2009we, Arkani-Hamed:2015bza} 
\begin{align}
    \frac{\left<\mathcal{R}\mathcal{R}\mathcal{R}\right>}{\left<\mathcal{R}\mathcal{R}\right>_\text{long}\,\left<\mathcal{R}\mathcal{R}\right>_\text{short}}
    \sim \epsilon_V \,g\,\left(\frac{k_\text{long}}{k_\text{short}}\right)^{\frac{3}{2} - \sqrt{\frac{9}{4}-\frac{C_{\Pi\Pi}}{H^2}}} \sim \epsilon_V \,g\,\left(\frac{k_\text{long}}{k_\text{short}}\right)^{1-\mathcal{O}(\epsilon_V)}\:.
    \label{eq:NG-estimate}
\end{align}
Here $\epsilon_V$ is the potential slow-roll parameter and $g$ is the coupling between the adiabatic fluctuations and the radion fluctuations. For a UV-localized inflaton, the interactions with the radion and the KK modes are Planck-suppressed making the signal apparently hard to observe (e.g. see ref.~\cite{Kumar:2018jxz} and possible ways around it~\cite{Kumar:2025anx}). 
However, the estimate in eq.~\eqref{eq:NG-estimate} is valid only in the regime in which $C_{\Pi\Pi}$ is approximately constant. This condition is satisfied only at late times during inflation. If we consider values of $k_\text{short}$ in eq.~\eqref{eq:NG-estimate} such that the corresponding mode exits the horizon when $C_{\Pi\Pi}$ is changing significantly, a different analysis is necessary to calculate the resulting signal. Figure~\ref{fig:masses-Crr-Css} shows $C_{\Pi\Pi}$ and $C_{\sigma\sigma}$ as a function of $c \Hsigma^2\eta^2$, for $\epsilon_V = 0.002, \eta_V = -0.014, c= 10^{-4}$. Also shown are $C_{rr}, C_{ss}$, the corresponding quantities for adiabatic and entropy fluctuations (see app.~\ref{app:EOM-fluctuations-along-adiabatic-entropy-direction}). We see that $C_{\Pi\Pi}$ is changing significantly at early times during inflation. It will be useful to calculate the effect of this time-dependence on the squeezed limit to see if this enhances the signal. Further, as we have discussed, modes with small enough $k$ experience only a few e-folds of inflation and may not be in the Bunch-Davies state, which can generate a signal in the folded triangle limit~\cite{Chen:2006nt} (see also ref.~\cite{Meerburg:2015yka}). All these discussions suggest a potentially interesting setup with signals that are strongly scale dependent. A detailed calculation is however necessary  to see whether the rich set of signals that are possible will be sufficiently large to detect. We will leave that for future work.

\section{Linear Fluctuations: Tensor Modes}
\label{sec:linear-fluctuations-tensor}

We next consider the tensor fluctuations in our setup. We calculate the power spectrum of the tensor modes, and recover the single-field expectation at late times (for large $k$ modes). We calculate the tensor tilt $n_t$ and show that it is $k$-dependent, unlike the single-field case, being strongly modified at large scales. 

The analysis for the tensor modes is simpler than for the scalar modes, but offers many insights that are obscured in the scalar case. The equations for the two polarizations of the tensor modes are decoupled (as in the single field case), and do not mix with the scalar modes at linear order, which simplifies the discussions considerably. The equations also admit an analytic solution, which helps to clarify the effect of the IR brane.

Including the tensor fluctuations $h_{ij}$, the metric to linear order is
\begin{align}
    \dd s^2 = a^2(\eta)\left(-\dd \eta^2 + \left(\delta_{ij} + h_{ij}\right)\dd x^i \dd x^j\right)\:.
\end{align}
Being symmetric, transverse, and traceless, there are two propagating degrees of freedom in $h_{ij}(x, \eta)$, as in the standard case. Since there is 3d rotational symmetry, we will work with the Fourier transform $h_{ij}(\bf{k}, \eta)$ in what follows, and suppress the $\bf{k}$ dependence. Expanding the Einstein equations to linear order, the $h_{ij}$ satisfy 
\begin{align}
    h_{ij}'' + 2 \frac{a'}{a} h_{ij}' + k^2 h_{ij} = 0\:.
\end{align}
It is more convenient to expand in terms of the polarization states $h_\lambda$ defined by
\begin{align}
    h_{ij} = \sum_{\lambda=+, \times} \epsilon_{ij}^\lambda h_\lambda\:,
\end{align}
where $\epsilon_{ij}^\lambda$ are the polarization tensors.

Defining the Mukhanov-Sasaki variable $f_\lambda = a(\eta) h_\lambda$, the function $f_\lambda$ satisfies
\begin{align}
    f_\lambda'' + \left(k^2-\frac{a''}{a}\right)f_\lambda = 0\:.
    \label{eq:MS-eqn-tensor}
\end{align}
To calculate $a''/a$, we use the Einstein equations in~\eqref{eq:EE-bkg} and work in the limit of $c \Hsigma^2\eta^2 \ll 1$. Defining $\mathcal{H} = a'/a$, the Einstein equations can be written as
\begin{align}
   -\frac{\dot{H}}{H^2} = 1-\frac{\mathcal{H}'}{\mathcal{H}^2} = \epsilon_V + 3 c \Hsigma^2\eta^2\:,
\end{align}
where we have used the background values for $\dot{\sigma}, \dot{\Pi}$ defined in eqns.~\eqref{eq:bkg-eqn-pi}, \eqref{eq:bkg-eqn-sigma}. Integrating this gives
\begin{align}
    \mathcal{H} = -\frac{1}{\eta}\left(1+\epsilon_V + c \Hsigma^2\eta^2\right)\:.
\end{align}
Note that here we have taken $\Hsigma$ to be approximately constant and expanded to linear order in $c, \epsilon_V$. In that limit, the above equation can be further integrated
around $\eta=\eta_*$ to give 
\begin{align}
    a(\eta) = -\frac{1}{\Hsigma\eta\,(\eta/\eta_*)^{\epsilon_V}}e^{-\frac12 c \Hsigma^{2}\eta^2}\:.
\end{align}
Here $\Hsigma, \epsilon_V$ are understood to be constant in a neighborhood of $\eta=\eta_*$. Note that the $c$ dependence matches that in eq.~\eqref{eq:scaleFactor}, up to linear order in $c \Hsigma^{2}\eta^2$, which is the order to which we are working.

Evaluating $a''/a$ and substituting in eq.~\eqref{eq:MS-eqn-tensor} we get
\begin{align}
    f_\lambda'' + \left(k^2-\frac{2+3\epsilon_V + c \Hsigma^{2}\eta^2}{\eta^2}\right)f_\lambda = 0\:.
\end{align}
While this equation can be solved exactly, and we will do that, we can already understand the effect of $c$ on the mode functions and the power spectrum. Heuristically, we can use the standard single-field result if we replace
\begin{align}
    \epsilon_V \to \epsilon_V + \frac{1}{3}c \Hsigma^{2}\eta_\text{HC}^2(k)\:,
    \label{eq:power-spectrum-tensor-approximation}
\end{align}
where $\eta_\text{HC}$ is to be evaluated at Hubble crossing and is a function of $k$, as indicated above. Using the result $n_t = -2\epsilon_V$ from the single-field case, we expect the tilt here to be
\begin{align}
    n_t = -2\epsilon_V - \frac{2}{3}c \Hsigma^{2}\eta_\text{HC}^2(k)\:.
\end{align}
This suggests that the power spectrum will be more red-tilted at smaller $k$ values, which is where the effect of a non-zero $c$ starts to show up. However this estimate is only expected to hold for $k$ values such that the shift in $\epsilon_V$ in eq.~\eqref{eq:power-spectrum-tensor-approximation} is at most of the same order as $\epsilon_V$ itself.

As the effect of $c$ was absorbed in $\epsilon_V$, and since the amplitude of the tensor modes is independent of $\epsilon_V$ in the single-field case, we expect that to also hold here. Using the amplitude for the scalar power spectrum from eq.~\eqref{eq:Delta-S-large-k} we get
\begin{align}
    r = 16\epsilon_V + 48 c \Hsigma^{2}\eta_\text{HC}^2(k)\:.
\end{align}
This suggests that we expect an enhancement of gravitational waves at small $k$ values. 
We now quantify this estimate by solving the equation for $f_\lambda$ exactly. 

On closer inspection, the equation for $f_\lambda$ can be rewritten as
\begin{align}
    f_\lambda'' + \left(\widetilde{k}^2-\frac{2+3\epsilon_V}{\eta^2}\right)f_\lambda = 0\:,\:\: \widetilde{k}^2 = k^2 - c \Hsigma^2\:,
\end{align}
and has the general solution
\begin{align}
    f_\lambda (k, \eta) = \frac{\sqrt{\pi}}{2}\sqrt{-\eta} \Bigg(A(k)\, H_\nu^{(1)}(-\widetilde{k}\eta) + B(k)\,H_\nu^{(2)}(-\widetilde{k}\eta)\Bigg)\:,\:\: \nu = 3/2 + \epsilon_V\:,
\end{align}
where $A(k),B(k)$ are $k$ dependent integration constants, to be determined by the initial conditions,  and $H^{(1)}_\nu(H^{(2)}_\nu)$ is the Hankel function of first (second) kind, of order $\nu$. 

From the general solution, there are a couple of observations one can make. The first thing to note is that for $k \lesssim \sqrt{c} H_\sigma$, $\widetilde{k}$ is imaginary, which seems to indicate an instability. 
This is related to the fact that small $k$ modes never experience any period of inflation (e.g. see fig.~\ref{fig:comovingHubbleRadius}). This is the effect of the IR brane on the dynamics and we see it clearly in the equation for $f_\lambda$ here. The same effect is also present for the scalar fluctuations, but is obscured due to the coupled nature of the equations and the lack of an analytical solution for all $k$. 
The second thing to note is that the effect of a non-zero $c$ is invisible for $\eta\to0$, since the term $\propto 1/\eta^2$ dominates. This is the late time or large $k$ limit, where we recover the single-field result. This clarifies that the effect from the IR brane is an ``IR'' effect, not seen at $\eta\to0$ (which is the UV) but only for larger $\eta$. This also makes sense from the 5D picture since the IR brane is quite far from the UV brane towards the end of inflation, $\eta\to0$. Finally, the presence of the IR brane leads to a non-zero $B$ (given the initial condition in eq.~\eqref{eq:mode-function-tensor-initial-condition}) and excites a non Bunch-Davies like state (in terms of $\widetilde{k}$).  

The $k$ dependent constants $A(k),B(k)$ can be solved by imposing the initial condition for $f_\lambda$ to be a plane wave at $\eta = \eta_i$:
\begin{align}
    f_\lambda(k, \eta_i) = \frac{1}{\sqrt{2k}}e^{-ik\eta_i}\:, f'_\lambda(k,\eta_i) = -i \sqrt{\frac{k}{2}}  e^{-ik\eta_i}\:.
    \label{eq:mode-function-tensor-initial-condition}
\end{align}
The explicit expressions for $A, B$ are straightforward to calculate and do not give much insight. However it is useful to understand their behavior for large $k$ modes, i.e. in the limit $k\eta_i \to -\infty$. In this limit, $k = \widetilde{k}$  and an explicit computation gives
\begin{align}
    A(k) \:\:\underset{-k\eta_i\to\infty}{=}\:\: - e^{i\pi\epsilon_V/2}\:,\qquad B(k) \:\:\underset{-k\eta_i\to\infty}{=}\:\:0\:. 
    \label{eq:A-large-k-limit}
\end{align}

Having calculated the mode function, the dimensionless power spectrum of the tensor modes is straightforward to obtain and is given by
\begin{align}
    \Delta_h^2(k^2) &= \lim_{\eta\to 0} \:\: \frac{8}{M_4^2}\frac{k^3}{2\pi^2}
    \frac{1}{a^2(\eta)}|f_\lambda(k, \eta)|^2 
    \nonumber \\
    &= \lim_{\eta\to 0} \frac{8}{M_4^2}\frac{k^3}{2\pi^2}\frac{1}{a^2(\eta)}\frac{\pi}{4}(-\eta) \left|A(k)\, H_{3/2+\epsilon_V}^{(1)}(-\widetilde{k}\eta)+B(k)\, H_{3/2+\epsilon_V}^{(2)}(-\widetilde{k}\eta)\right|^2\:.
\end{align}
Using the late-time limit of Hankel functions
\begin{align}
    H_{\nu}^{(1)}(-\widetilde{\kappa}\eta)\underset{-\widetilde{\kappa}\eta \to 0}{ = }\:\frac{\Gamma(\nu)}{i\pi}\left(\frac{-\widetilde{\kappa}\eta}{2}\right)^{-\nu}\:,\:\:
    H_{\nu}^{(2)}(-\widetilde{\kappa}\eta)\underset{-\widetilde{\kappa}\eta \to 0}{ = }\:-\frac{\Gamma(\nu)}{i\pi}\left(\frac{-\widetilde{\kappa}\eta}{2}\right)^{-\nu}\:,
\end{align}
we get
\begin{align}
    \Delta_h^2(k^2) 
    &= \lim_{\eta\to 0} \frac{8}{M_4^2}\frac{k^3}{2\pi^2}\frac{1}{a^2(\eta)}\frac{\pi}{4}(-\eta) \,\Big|A(k)-B(k)\Big|^2 \,\frac{\Gamma^2(3/2+\epsilon_V)}{\pi^2} \left(-\frac{\widetilde{k}\eta}{2}\right)^{-3-2\epsilon_V}  \nonumber \\
    &=
    \frac{2}{\pi^2}\left(\frac{\Hsigma}{M_4}\right)^2
    \Big|A(k)-B(k)\Big|^2
    \left(\frac{\sqrt{k^2 - c \Hsigma^2}}{k_\text{ref}}\right)^{-2\epsilon_V}\left(\frac{k}{\sqrt{k^2 - c \Hsigma^2}}\right)^3\:,
    \label{eq:power-spectrum-tensor-exact}
\end{align}
where in the last line we have defined a reference momentum $k_\text{ref} = 1/\eta_*$. 

We know that at late times during inflation, the dynamics is that of a single-field model, and since large $k$ modes exit the horizon at late times, we expect to recover the power spectrum for a single field case. In the limit of large $k$, $k=\widetilde{k}$ and from eq.~\eqref{eq:A-large-k-limit}, $\left|A-B\right|^2 = 1$. In this limit we therefore get
\begin{align}
    \Delta_h^2(k^2) =
    \frac{2}{\pi^2}\left(\frac{\Hsigma}{M_4}\right)^2\left(\frac{k}{k_\text{ref}}\right)^{-2\epsilon_V}\:,
    \label{eq:power-spectrum-tensor-SFSR}
\end{align}
which is the expected result: a constant tilt of $-2\epsilon$ and a constant amplitude.

\begin{figure}[h]
\centering
\includegraphics[scale=0.65]{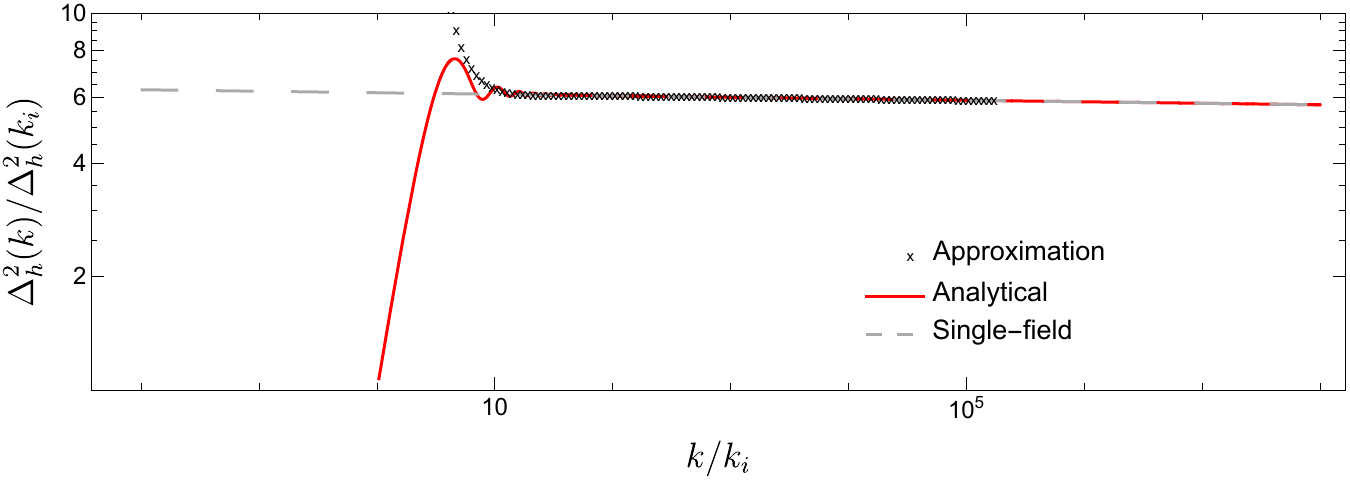}
\caption{\small{Rescaled power spectrum for the tensor fluctuations as a function of $k/k_i$. The approximation (in black cross) matches the analytical result (in red solid), and for large $k$, both of them match the slope expected from the late time single-field limit. We have taken $\epsilon_V = 0.002, \eta_V = -0.014, c = 10^{-4}$ and $x_i = -1/\sqrt{3c}$.}}
\label{fig:power-spectrum-tensor}
\end{figure}

The general expression for $\Delta_h$ shows clearly that both the amplitude and the tilt are modified due to the presence of the IR brane. Fig.~\ref{fig:power-spectrum-tensor} shows the power spectrum of the tensor fluctuations as a function of $k$. Shown are the analytical result (eq.~\eqref{eq:power-spectrum-tensor-exact}, in red solid), the result obtained from using the approximation (eq.~\eqref{eq:power-spectrum-tensor-approximation}, in black crosses), and the single-field expectation (eq.~\eqref{eq:power-spectrum-tensor-SFSR}, in gray dashed). We have rescaled the power spectrum with its value at the smallest $k$, denoted by $k_i$, and presented it as a function of $k/k_i$. We see that for large $k$, the slope of the analytical and approximate result for $\Delta_h^2$ match the single-field expectation well. The analytical result shows oscillations at small $k$, and the envelope is captured well by the approximate expression for $\Delta_h^2$ obtained by adjusting $\epsilon_V$ to a $k$ dependent value (eq.~\eqref{eq:power-spectrum-tensor-approximation}).

In summary, due to the presence of additional radion dynamics, there are two differences in the tensor power spectrum compared to the single field result:
\begin{itemize}
    \item Amplitude: the power spectrum is modified and shows oscillations at smaller $k$ values.
    \item Tilt: the tilt is not constant, and is significantly modified at low $k$, showing oscillations.
\end{itemize}
Both the effects, seen at large scales, come from the early periods of inflation when the curvature perturbation is generated by the radion.

\begin{figure}[h]
\centering
\includegraphics[scale=0.7]{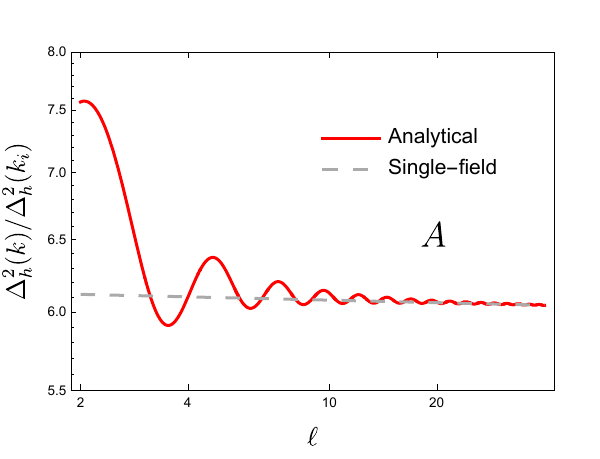}
\qquad
\includegraphics[scale=0.7]{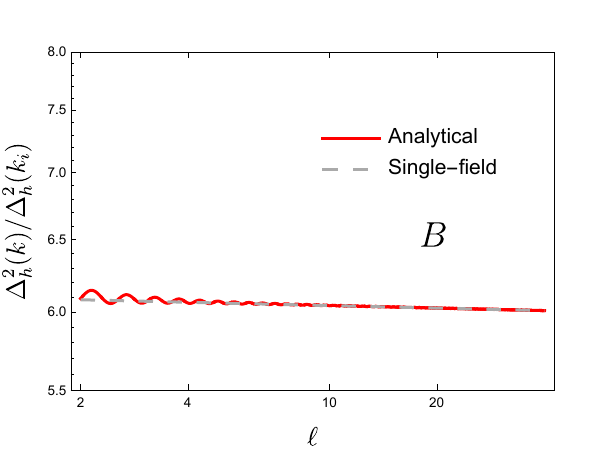}
\caption{\small{Rescaled power spectrum for the tensor fluctuations as a function of $\ell$ for two locations of the CMB window $A, B$ (see discussion near fig.~\ref{fig:power-spectrum-adiabatic-slidingCMBwindow}).  
Here $\epsilon_V = 0.002, \eta_V = -0.014, c = 10^{-4}$ and $x_i = -1/\sqrt{3c}$.}}
\label{fig:power-spectrum-tensor-AB}
\end{figure}

Here we quantify how the deviations in tensor power spectrum from the single-field case scale with CMB angular mode $\ell$. Figure~\ref{fig:power-spectrum-tensor-AB} shows the rescaled tensor power spectrum as a function of $\ell$ for two benchmark location of CMB window (corresponding to $A$ and $B$ in fig.~\ref{fig:power-spectrum-adiabatic-slidingCMBwindow}). We see that even for the most optimistic scenario, the deviations from single-field case are below a percent level after $\ell \gtrsim 20$ and the largest deviations from the single-field scenario occur at small $\ell$, where galactic dust and cosmic variance would complicate detection. Future improvements in  low $\ell$ measurements would be necessary to observe the deviations coming from the presence of the radion.

\section{Summary and Conclusion}
\label{sec:conclusion}

In this work we have considered a time-dependent 5D geometry that provides an interesting candidate for the inflationary phase of the early Universe. We considered a spacetime with a warped fifth dimension between a UV and an IR brane in which the UV brane tension is detuned due to a localized inflaton potential. This potential, assumed to be approximately flat so that the inflaton slow-rolls, causes the 5D geometry to be time-dependent. The fifth dimension grows during inflation, with results that asymptote to conventional exponential inflationary growth at late times.

Observing that the motion of the brane leads to a time-dependent volume and therefore a time-dependent 4D Planck scale, we constructed a 4D theory that matches the background solution. This 4D theory has two light scalars, an inflaton and a radion, both of which have a time-dependent background and contribute to the evolution of the background geometry of the Universe. The resulting scale factor of the 4D FLRW geometry showed deviations from a slow-roll de Sitter expectation at early times. At early enough time, the equation of state parameter $w$ for the fluid sourcing the geometry is larger than $-1/3$. 
A novel aspect of the dynamics is therefore that there is a starting point of inflation, which corresponds to the branes being close together. 
This means that a given $k$ mode undergoes only a finite period of inflation (e.g. see fig.~\ref{fig:comovingHubbleRadius}). In fact, sufficiently small $k$ modes never experience inflation at all. 

This setup allows us  to investigate the dynamics near the starting point of inflation. As we saw, the adiabatic fluctuations near the start of inflation come from the radion rather than the inflaton, causing a visible reduction in the power spectrum for low $\ell$ CMB modes. 
The finite time period of inflation also introduces dependence of the power spectrum on the initial conditions, as low-$\ell$ modes may have experienced only a few e-folds of inflation. For highly inhomogeneous initial conditions this can introduce further deviations from the usual slow-roll expectation on large scales.

The adiabatic scalar power spectrum has a blue tilt at large scales due to the radion dynamics, and a red tilt at small scales consistent with the single-field expectation. 
The CMB temperature fluctuations, which track the adiabatic fluctuations, show large error bars at low $\ell$ values, coming from cosmic variance as well as other experimental details. 
There are indications that a slight blue tilt at low $\ell$ values will fit the CMB data better than the single field expectation~\cite{Petretti:2024mjy}. This is promising given that the dynamics we have considered generates this blue tilt in a very robust manner, without explicitly putting it in. 
Nonetheless, the spectrum cannot deviate too much from the usual single field expectation, as it would come into conflict with current CMB measurements. Whether the modes corresponding to the deviation in the power spectrum are seen in the CMB or not depends on the number of e-folds of inflation that occur after they leave the horizon and the subsequent post-inflationary cosmology. In our analysis the time at which inflation ends, $\eta = \eta_f$, is left as a free parameter, and for a range of $\eta_f$ values the low-$\ell$ modes of the CMB can show a blue-tilt, while the well-measured modes for $\ell \gtrsim 20$ follow the usual inflationary prediction.

Apart from the adiabatic mode, we also considered the isocurvature modes in this scenario, and showed that they decay once they become superhorizon, hence not leaving any observational imprints on late time cosmology.  We also calculated the power spectrum of the tensor modes and showed that the dynamics leads to oscillatory features at large scales, as seen in fig.~\ref{fig:power-spectrum-tensor}. While we have not seen any tensor modes yet, the prediction of the theory is that the generated tensor modes would have a varying tilt and would change from a constant red-tilt at small scales to an oscillatory tilt at larger scales, potentially showing enhancement at small scales. However, as we estimated, even for the most optimistic scenario, the effects are negligible for $\ell \gtrsim 20$ and would be challenging to observe. Finally, we briefly discussed how the early time dynamics during inflation, whose imprints are at large scales, can modify primordial non-Gaussianities, e.g. in the squeezed limit of the three point function.

It is exciting that none of these features are engineered. They are natural consequences of the finite starting time and the dominance of the radion early on.
Uncertainties in predictions due to unknown classical fluctuations and uncertainties in the initial vacuum state, mean that specific results can change. 
The important point is that the low $\ell$ CMB modes can contain information about the initial conditions for inflation, and our model gives one possible scenario. There are additional scenarios one can consider where inflation is preceded by a different epoch, which sets the initial conditions. Given that the data for low $\ell$ modes show deviations from a single-field result (but is consistent within the error bars), this work serves as motivation to further pin down this elusive CMB regime.

There are several future natural directions to consider. While we derived a 4D theory by matching the background solution, it would be useful to derive it from a dimensional reduction. It would be important to identify the masses of the massive spin-2 KK modes and their signatures in the non-Gaussianities and their possible cosmological collider signals. One can also consider a more general scenario where both UV and IR brane tensions are detuned~\cite{Mishra:2022fic}, which can lead to a more rich set of possibilities. Most importantly, it would be important to perform a detailed fit to the CMB and LSS data.

\subsection*{Acknowledgements}
We would like to  
thank A. Bodas, P. Chakraborty, P. Du, M. Ekhterachian, M. Fasiello, S. Kumar, M. Muenchmeyer, S. Paban, H. Peiris and A. Thavanesan for discussions and useful comments. RKM would like to thank KITP, MITP and CERN for their hospitality, where part of this work was completed. This research was supported in part by grant NSF PHY-2309135 to the Kavli Institute for Theoretical Physics (KITP). RKM is grateful to the Mainz Institute for Theoretical Physics (MITP) of the Cluster of Excellence PRISMA+ (Project ID 390831469), for its hospitality and its partial support during the completion of this work. MN is supported by NSF Award PHY-2310717. We would like to acknowledge GRASP Initiative funding provided by Harvard University.

\appendix

\section{5D Picture in de Sitter Slicing}

\label{app:dS-slicing}

When discussing the 5D picture, we worked in coordinates where constant-$z$ slices of the extra dimensions were described by a Minkowkski metric. The 5D geometry consisted of a UV brane which is moving due to a detuned tension and a static IR brane. In the Minkowski slicing it seems that there should be a KK tower with masses set by the IR scale; in principle this tower can be lighter than the scale set by $H$. If so, this would result in a large number of light modes in the spectrum that would need to be included in any consistent effective theory. However, studies of the RSII model in inflation where the IR brane is effectively at infinite distance, $z_\ir \to \infty$ (where the KK scale is effectively zero), show a gapped spectrum with the lightest modes having Hubble-scale masses~\cite{Garriga:1999bq, Karch:2000ct, Kobayashi:2000yh, Gabadadze:2021dnk, Hubisz:2024xnj}. In this case the mass gap is set by the Hubble scale instead.

In this appendix we clarify the situation for our setup.  
We consider the 5D geometry in coordinates where the 4D slices each have a de Sitter metric (ignoring slow-roll corrections). In these coordinates the UV brane is static, while the IR brane moves~\cite{Karch:2020iit}. In the de Sitter slicing we now consider, the IR brane approaches the horizon only at infinite time whereas in the flat slicing, the IR brane crosses the horizon at finite time. In either frame, we have a radion whose mass saturates at order the Hubble scale. 

We start with the Minkowski coordinates~\eqref{eq:5D_metric-general}, and transform to coordinates where the metric is given by~\cite{DeWolfe:1999cp}
\begin{align}
    ds^2 &= (H L)^2 \sinh^2 \left( \frac{r_h - r}{L} \right) ds_4^2 + dr^2 \, , \nonumber
    \\
    ds_4^2 &= - d\tau_{\rm dS}^2 + e^{2H\tau_{\rm dS}} d\vec x . \, d\vec x \, ,
    \label{eq:metric_ds}
\end{align}
where $H$ is the (constant) hubble scale, $L$ is the AdS radius, $r$ is the extra-dimensional coordinate, and $r_h$ is the location of the horizon in the de Sitter slicing. The AdS boundary is at $r=-\infty$ and the coordinates are restricted so that $r < r_h$. 

The Minkowski coordinates $(z, \tau)$ are related to the de Sitter coordinates $(r, \tau_{\rm dS})$ by the transformations
\begin{align}
    \tau - \frac{z_0}{HL}&= - \frac{1}{H} \coth \left( \frac{r_h - r}{L} \right) e^{-H \tau_{\rm dS}}  \, , \nonumber
    \\
    z &= \frac{1}{H} \csch \left( \frac{r_h - r}{L} \right)   e^{-H \tau_{\rm dS}} \, .
    \label{eq:ds_coords}
\end{align}
The LHS in the first line of eq.~\eqref{eq:ds_coords} is chosen to ensure that the UV brane is static in the new coordinates. Note that the conformal time $\eta$ is given by $\eta =  \tau - z_0/(HL)$ (see discussions near eq.~\eqref{eq:tau-eta-relation}), and is the same as the LHS of the first equation in~\eqref{eq:ds_coords}. Since $\tau$ takes values in the range $(0, \infty)$, the above equation fixes the range of $\tau_\text{dS}$. The initial value for $\tau_\text{dS}$ is obtained by setting $\tau=0$ and $r = r_\text{uv}$ in the first equation.

Given the change of coordinates, we can transform the solutions $z_\uv = z_0 - H L \, \tau$ and $z_\ir = const$ to the de Sitter coordinates, and find the corresponding motion of the branes. The result is that the UV brane is static at $r = r_\uv$, where
\begin{align}
    r_\uv = r_h - L\arcsech(H L) \, .
\end{align}
In the de Sitter sliced coordinates the IR brane moves according to 
\begin{align}
    r_\ir (\tau_{\rm dS}) = r_h - L \arcsinh \left(\frac{e^{-H\tau_{\rm dS}}}{Hz_\ir} \right) \, ,
\end{align}
and asymptotically approaches the horizon $r=r_h$ as $\tau_{\rm dS} \to \infty$. 
This happens at a finite $\eta$, as can be seen by inverting eq.~\eqref{eq:ds_coords}.

\section{Detailed Expressions for the Fluctuation Equations}
\label{app:EOM-fluctuations}
This appendix collects the expressions that arise when studying fluctuations. The discussion is taken from ref.~\cite{Lalak:2007vi} and modified to our case.
\subsection{Fluctuations along field directions}
\label{app:EOM-fluctuations-along-field-direction}
We present here the expression for various terms appearing in the equations of motion for the gauge invariant variables $Q_\Pi, Q_\sigma$. The equations are
\begin{align}
    \ddot Q_\Pi + \left(3H+B_{\Pi\Pi}\right) \dot{Q_\Pi} + B_{\Pi\sigma}\dot Q_\sigma + \left(\frac{k^2}{a^2} + C_{\Pi\Pi}\right)Q_\Pi + C_{\Pi\sigma} Q_\sigma &= 0\:,
    \\
    \ddot Q_\sigma + \left(3H+B_{\sigma\sigma}\right) \dot{Q_\sigma} + B_{\sigma\Pi}\dot Q_\Pi + \left(\frac{k^2}{a^2} + C_{\sigma\sigma}\right)Q_\sigma + C_{\sigma\Pi} Q_\Pi &= 0\:,
\end{align}
where
\begin{align}
    B_{\Pi\Pi} &= 0\:, \\
    B_{\sigma\sigma} &= \dot{\Pi}\,\partial_\Pi \log f(\Pi)\:, \\
    B_{\Pi\sigma} &= -\dot{\sigma}\,\partial_\Pi f(\Pi)\:, \\
    B_{\sigma\Pi} &= \dot{\sigma}\,\partial_\Pi \log f(\Pi)\:,
\end{align}
and
\begin{align}
    C_{\Pi\Pi} 
    &= 
    \partial^2_\Pi\,V 
    + 
    \frac{2\dot{\Pi} \, \partial_\Pi\,V}{M_4^2\,H}
    + \frac{3\,\dot{\Pi}^2}{M_4^2}
    - \frac{\dot{\Pi}^4}{2 M_4^4\, H^2}
    - \frac12f(\Pi)(\partial_\Pi\,\log f)^2\dot{\sigma}^2
    \nonumber \\
    &\qquad
    - \frac12 f(\Pi)(\partial_\Pi^2 \log f)\,\dot{\sigma}^2
    - \frac{f(\Pi)\dot{\Pi}^2\dot{\sigma}^2}{2 M_4^4\,H^2}
    \:, \\
    C_{\sigma\sigma} 
    &= 
    \frac{1}{f(\Pi)}\partial_\sigma^2 V
    + \frac{2\dot{\sigma}\partial_\sigma V}{M_4^2\,H}
    - \frac{f(\Pi)\dot{\sigma}^2\dot{\Pi}^2}{2 M_4^4\,H^2}
    - \frac{f^2(\Pi)\dot{\sigma}^4}{2 M_4^4\,H^2}
    + \frac{3 f(\Pi)\dot{\sigma}^2}{M_4^2}
    \:, \\
    C_{\Pi\sigma} 
    &= 
    \partial_\Pi\partial_\sigma\,V
    + \frac{f(\Pi)\dot{\sigma}\partial_\Pi\,V}{M_4^2\,H}
    + \frac{\dot{\Pi}\partial_\sigma\,V}{M_4^2\,H}
    - \frac{f(\Pi)\dot{\Pi}^3\dot{\sigma}}{2\,M_4^4\,H^2}
    - \frac{f^2(\Pi)\dot{\Pi}\dot{\sigma}^3}{2\,M_4^4\,H^2}
    + \frac{3f(\Pi)\dot{\sigma}^2}{M_4^2}
    \:, \\
    C_{\sigma\Pi} 
    &= 
    \frac{1}{f(\Pi)}\partial_\Pi\partial_\sigma\,V
    + \frac{\dot{\sigma}\partial_\Pi\,V}{M_4^2\,H}
    + \frac{\dot{\Pi}\partial_\sigma\,V}{f(\Pi)M_4^2\,H}
    - \frac{1}{f(\Pi)} (\partial_\Pi \log f)\partial_\sigma\,V
    \nonumber \\
    &\qquad
    + (\partial_\Pi^2 \log f)\dot{\Pi}\dot{\sigma}
    - \frac{\dot{\Pi}^2\dot{\sigma}}{2\,M_4^4\,H^2}
    - \frac{f(\Pi)\dot{\Pi}\dot{\sigma}^3}{2\,M_4^2\,H^2}
    + \frac{3\dot{\Pi}\dot{\sigma}}{M_4^2}\:.
\end{align}
Plugging in the explicit form of $V(\Pi, \sigma), f(\Pi)$ from eq.~\eqref{eq:full-action} and using the background solutions, to leading order in the potential slow-roll parameters, these quantities are given as
\begin{align}
    \frac{B_{\sigma\sigma}}{H^2} 
    &=
    2 c \Hsigma^2\eta^2 \left(-\frac{3 \epsilon_V  \left(c \Hsigma^2\eta^2-1\right)^3}{\left(2 c \Hsigma^2\eta^2-3\right) \left(4 c \Hsigma^2\eta^2+1\right)}-1\right)
    \:,\:\:  \frac{B_{\Pi\Pi}}{H^2} = 0 \\
    \frac{B_{\sigma\Pi}}{H^2} &=
    -\frac{2 \sqrt{3c \Hsigma^2\eta^2}  \sqrt{\epsilon_V } \left(c \Hsigma^2\eta^2-1\right)^2}{2 c \Hsigma^2\eta^2-3}
    \:,\:\:
    \frac{B_{\Pi\sigma}}{H^2} =
    -\frac{2 \sqrt{3c \Hsigma^2\eta^2}  \sqrt{\epsilon_V } \left(c \Hsigma^2\eta^2-1\right)}{2 c \Hsigma^2\eta^2-3}
    \:.
\end{align}
\begin{align}
    \frac{C_{\Pi\Pi}}{H^2}
    &=
    -\frac{3 \epsilon_V  \left(132 c^3 \Hsigma^6\eta^6-338 c^2 \Hsigma^4\eta^4+77 c \Hsigma^2\eta^2-7\right) \left(c \Hsigma^2\eta^2-1\right)^3}{\left(3-2 c \Hsigma^2\eta^2\right)^2 \left(4 c \Hsigma^2\eta^2+1\right)}-2 c \Hsigma^2\eta^2+2
    \:, \\
    \frac{C_{\sigma\sigma}}{H^2}
    &=
    3 \left(c \Hsigma^2\eta^2-1\right)^2 \left(-\frac{6 \epsilon_V  \left(c \Hsigma^2\eta^2-3\right) \left(c \Hsigma^2\eta^2-1\right)}{\left(3-2 c \Hsigma^2\eta^2\right)^2}+\eta_V \right)
    \:,\\
    \frac{C_{\Pi\sigma}}{H^2}
    &=
    \frac{2 \sqrt{3} \left(c \Hsigma^2\eta^2\right)^{3/2} \sqrt{\epsilon_V } \left(c \Hsigma^2\eta^2-1\right)}{2 c \Hsigma^2\eta^2-3}
    \:,\\
    \frac{C_{\sigma\Pi}}{H^2}
    &=
    -\frac{4 \sqrt{3c \Hsigma^2\eta^2}  \sqrt{\epsilon_V } \left(c \Hsigma^2\eta^2-1\right)^2}{2 c \Hsigma^2\eta^2-3}
    \:.
\end{align}

\begin{figure}[h]
\centering
\includegraphics[scale=1]{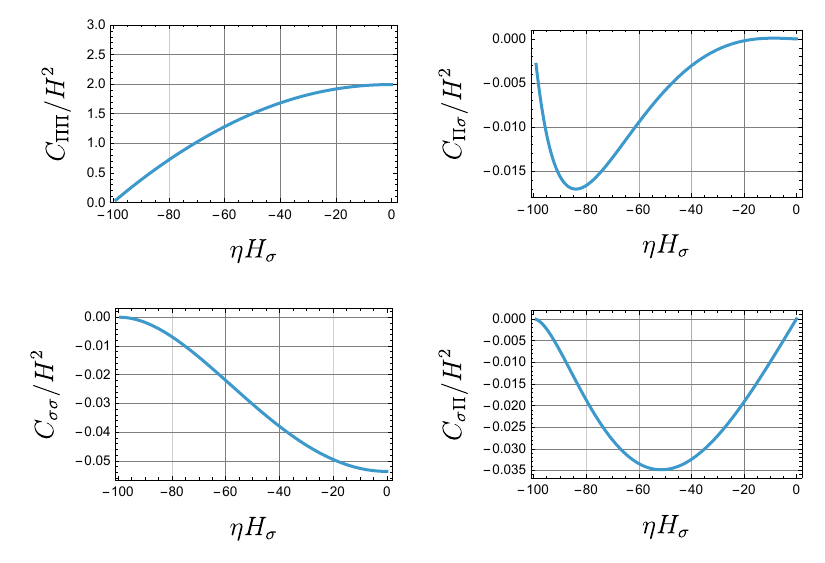}
\\
\includegraphics[scale=1]
{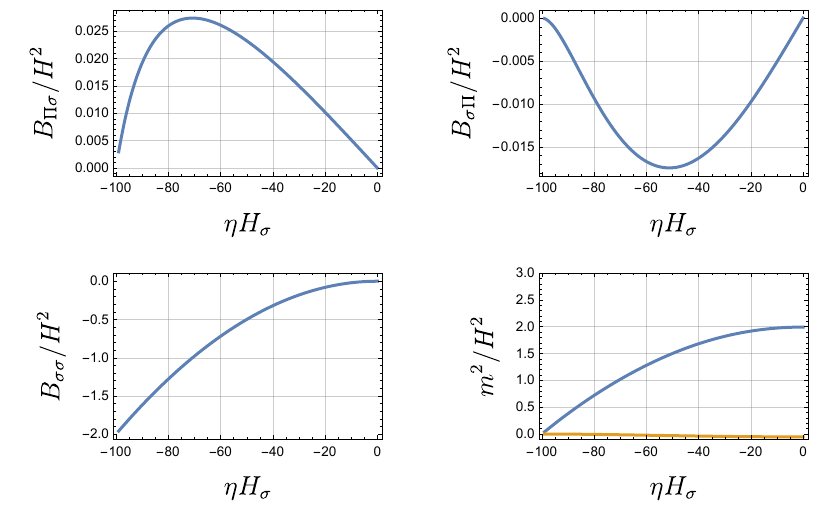}
\caption{\small{$B_{ij}, C_{ij}$ as a function of $\eta \Hsigma$ for $\epsilon_V = 0.002, \eta = -0.014, c= 10^{-4}$. Bottom right shows the eigenvalues of the $C_{ij}$ matrix.}}
\label{fig:MassMatrixOfFluctuations}
\end{figure}

Figure~\ref{fig:MassMatrixOfFluctuations} shows $B_{ij}, C_{ij}$ for a representative choice of $\epsilon_V, \eta_V$. Further, given the mass matrix elements $C_{ij}$, we can diagonalize the mass matrix, to get the two mass eigenvalues (ignoring the mixed Hubble friction terms). These are also shown in fig.~\ref{fig:MassMatrixOfFluctuations} (bottom right).

\subsection{Fluctuations along adiabatic/entropy directions}
\label{app:EOM-fluctuations-along-adiabatic-entropy-direction}
To solve the equations for fluctuations, it is useful to rotate to the adiabatic and entropy directions as done in sec.~\ref{subsec:adiabatic-entropy-fluctuations}. The quantities $b_{rr}, b_{rs}, b_{sr}, b_{ss}, c_{rr}, c_{ss}, c_{rs}, c_{sr}$ appearing in eq.~\eqref{eq:qr-qs-eqn} are given as
\begin{align}
    b_{rr} &=  \frac{a}{\Hsigma} B_{rr} - a \frac{H}{\Hsigma}\:,\:\:
    b_{ss} =  \frac{a}{\Hsigma} B_{ss} - a \frac{H}{\Hsigma}\:,\:\:
    b_{rs} = \frac{a}{\Hsigma}B_{rs}\:,\:\:b_{sr} = \frac{a}{\Hsigma}B_{sr}
    \nonumber \\
    c_{rr} &= \frac{a^2}{\Hsigma^2}C_{rr}\:,\:\: c_{ss} = \frac{a^2}{\Hsigma^2}C_{ss}\:,\:\:
    c_{rs} = \frac{a^2}{\Hsigma^2}C_{rs}\:,\:\: c_{sr} = \frac{a^2}{\Hsigma^2}C_{sr}\:.
\end{align}
The $B_{ij}, C_{ij}$ are given as
\begin{align}
    B_{rr} &= B_{ss} = 3H\:,\:\: B_{rs} = -B_{sr} = \frac{2V_s}{\dot{X}}\:,
\end{align}
\begin{align}
    C_{rr} &= -\frac12 \partial_\Pi (\log f) \left(V_r \sin^2\theta  \cos \theta +V_s \sin \theta \left(\cos^2\theta+1\right)\right)
    -\frac{\dot{X}^4}{2 H^2 M_4^4}+\frac{2 V_r \dot{X}}{H M_4^2}+\frac{3 \dot{X}^2}{M_4^2}+V_{rr}-\left(\frac{V_s}{\dot{X}}\right)^2\:,
    \nonumber \\
    C_{ss} &= \frac12 V_r \left(\sin^2\theta+1\right) \cos\theta \,\partial_\Pi(\log f)+\frac12 V_s \sin\theta \cos^2\theta\, \partial_\Pi(\log f)
    \nonumber\\
    &\qquad\qquad\qquad\qquad-\frac14 \dot{X}^2 \left(2\partial_\Pi^2(\log f)+\left(\partial_\Pi(\log f)\right)^2\right)-\left(\frac{V_s}{\dot{X}}\right)^2+V_{ss}\:,
\end{align}
\begin{align}
    C_{rs} &= \partial_\Pi(\log f) \left(V_r \sin^3\theta-V_s \cos^3\theta\right)+\frac{V_s \dot{X}}{H M_4^2}+\frac{6 H V_s}{\dot{X}}+\frac{2 V_r V_s}{\dot{X}^2}+2 V_{rs}\:,
    \nonumber \\
    C_{sr} &= \frac{V_s \dot{X}}{H M_4^2} - \frac{6 H V_s}{\dot{X}}-\frac{2 V_r V_s}{\dot{X}^2}\:.
\end{align}
where
\begin{align}
    V_r &= \cos \theta\, \partial_\Pi V(\Pi, \sigma) + \sin \theta\,\sech\left(\frac{\Pi}{\sqrt{6} M_4}\right) \partial_\sigma V(\Pi, \sigma)\:,
    \nonumber \\
    V_s &= -\sin \theta\, \partial_\Pi V(\Pi, \sigma) + \cos \theta\,\sech\left(\frac{\Pi}{\sqrt{6} M_4}\right) \partial_\sigma V(\Pi, \sigma)\:,
\end{align}
and
\begin{align}
    V_{rr} &= \cos^2 \theta\, \partial_\Pi^2 V(\Pi, \sigma) + \sin^2 \theta\,\sech^2\left(\frac{\Pi}{\sqrt{6} M_4}\right) \partial_\sigma^2 V(\Pi, \sigma) + 2\cos\theta\,\sin\theta\,\sech\left(\frac{\Pi}{\sqrt{6} M_4}\right)\partial_\Pi\partial_\sigma V(\Pi, \sigma)\:,
    \nonumber \\
    V_{ss} &= \sin^2 \theta\, \partial_\Pi^2 V(\Pi, \sigma) + \cos^2 \theta\,\sech^2\left(\frac{\Pi}{\sqrt{6} M_4}\right) \partial_\sigma^2 V(\Pi, \sigma) - 2\cos\theta\,\sin\theta\,\sech\left(\frac{\Pi}{\sqrt{6} M_4}\right)\partial_\Pi\partial_\sigma V(\Pi, \sigma)\:,
    \nonumber \\
    V_{rs} &= -\sin \theta\,\cos\theta\, \partial_\Pi^2 V(\Pi, \sigma) + \sin\theta\,\cos \theta\,\sech^2\left(\frac{\Pi}{\sqrt{6} M_4}\right) \partial_\sigma^2 V(\Pi, \sigma) 
    \nonumber \\
    &\qquad\qquad+ \left(\cos^2\theta-\sin^2\theta\right)\sech\left(\frac{\Pi}{\sqrt{6} M_4}\right)\partial_\Pi\partial_\sigma V(\Pi, \sigma)\:.
\end{align}

\section{Sensitivity to Initial Conditions}
\label{app:initial-conditions}
In the main text we obtained the power spectrum assuming the initial condition on the mode functions $q_r, q_s$ is given as in eq.~\eqref{eq:qr-qs-ic}. 
Here we show what happens when we deviate from that choice.

\begin{figure}[h]
\centering
\includegraphics[scale=0.51]{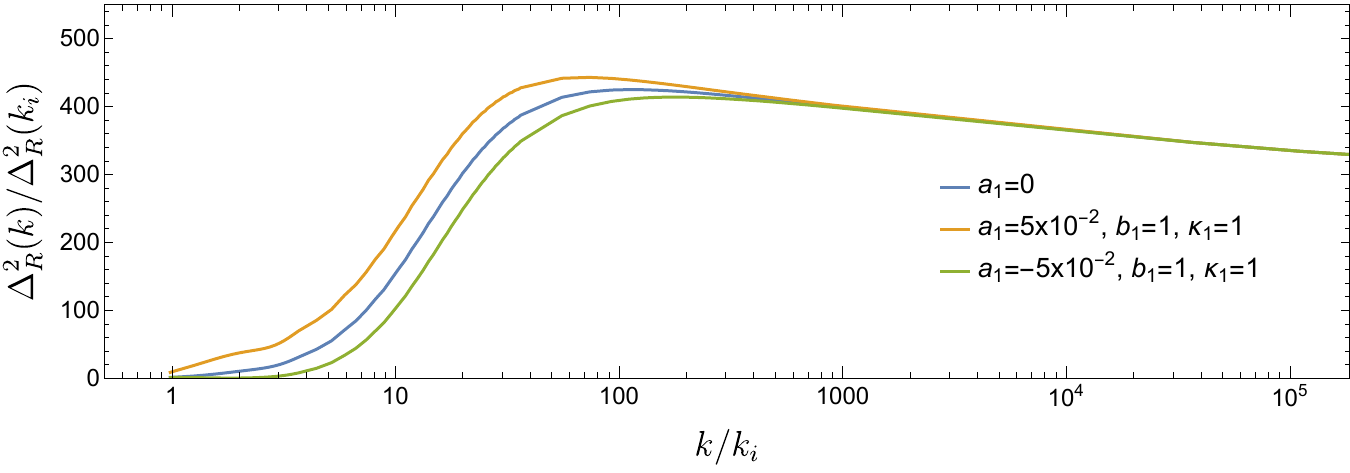}
\\
\includegraphics[scale=0.51]{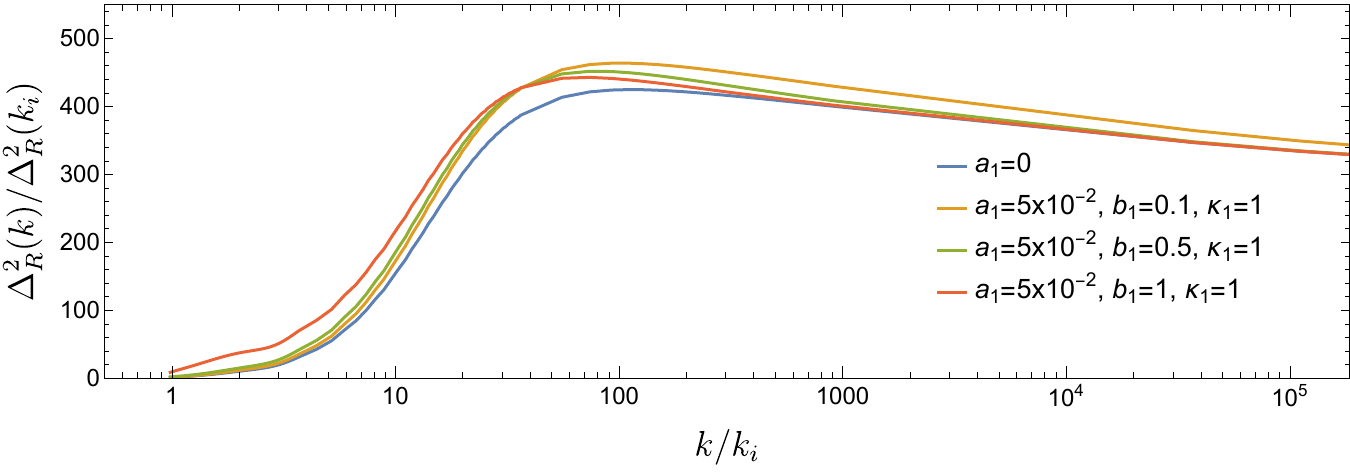}
\\
\includegraphics[scale=0.51]{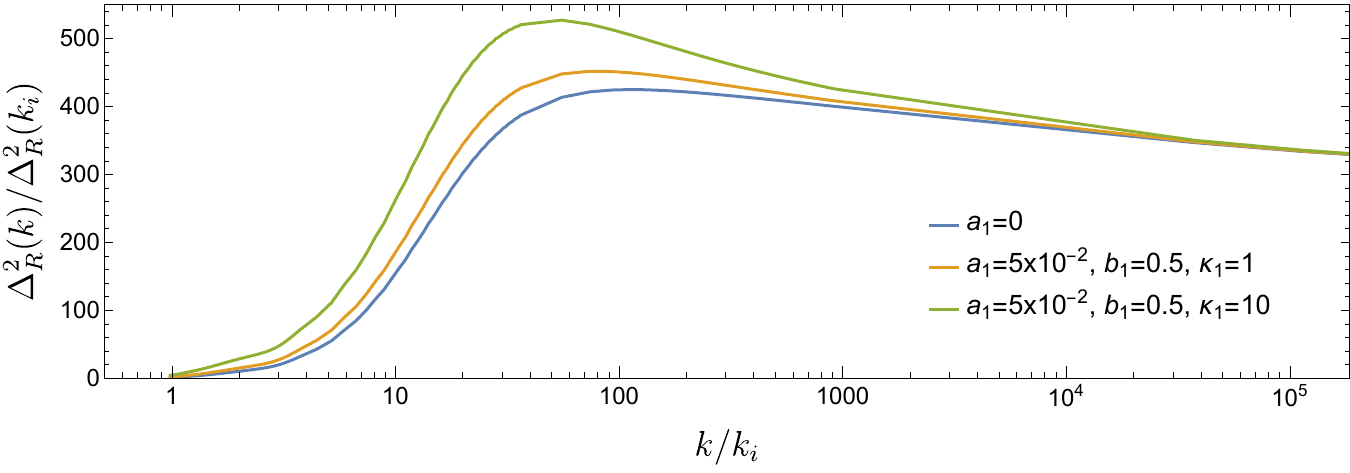}
\caption{\small{Effect of changing the initial conditions from a plane wave to that parameterized in eq.~\eqref{eq:non-BD-ic}.}}
\label{fig:PowerSpectrum-Adiabatic-vary-ic}
\end{figure}

The choice of a plane wave initial condition is justified for large $k$ modes (or equivalently modes for which $\eta$ can be taken to be large and negative), but it is less clear what is the unique choice for small $k$. To quantify that in a heuristic fashion, we will parameterize a modification to this assumption by taking
\begin{align}
    q_r(x_i) = \frac{A(\kappa)}{a(x_i)\sqrt{2\kappa}}e^{-i \kappa x_i}\:,\:\:
    q_s(x_i) = \frac{A(\kappa)}{a(x_i)\sqrt{2\kappa}}e^{-i \kappa x_i}\:,
    \:\:
    A(\kappa) = 1 + a_1 (\kappa_1/\kappa)^{b_1} + \cdots\:,
    \label{eq:non-BD-ic}
\end{align}
and choosing different choice of $a_1, b_1, \kappa_1$. For large $\kappa$, $A(\kappa)$ should reduce to $1$, which means $b_1>0$.  Figure~\ref{fig:PowerSpectrum-Adiabatic-vary-ic} shows the effect of changing $a_1, b_1, \kappa_1$ on the power spectrum. We see that the blue tilt at low $k$ value persists, at least for the choice of $a_1, b_1, \kappa_1$ we have used. In this we have imposed the initial condition at $x = x_i = -1/\sqrt{3c}$.

\newpage
\bibliographystyle{utphys}
\bibliography{refs}

\end{document}